\documentclass{aa}  

\usepackage{graphicx}
\usepackage{txfonts}

\usepackage{ulem}

\usepackage{xcolor}

\usepackage{soul}

\usepackage{float}

\usepackage{array, multirow, bigdelim, makecell, booktabs} 

\usepackage{threeparttable}

\usepackage{lscape}

\usepackage{longtable}
\usepackage[colorlinks = true, linkcolor=blue, citecolor = blue]{hyperref}
\newcommand{\numTotal}{192~}
\newcommand{\numGS}{80~}
\newcommand{\numSThree}{51~}
\newcommand{\numSThreeApproved}{45~} %2 additional from XMM/Chandra detections
\newcommand{\numSThreeApprovedP}{47~}
\newcommand{\numSThreeApprovedPdistance}{46~}

\newcommand{\numPB}{17~}
\newcommand{\numPorb}{190~}
\newcommand{\numPorbMdonor}{175~}
\newcommand{\numSpT}{38~}
\newcommand{\numGaia}{159~}
\newcommand{\numGaiaBJ}{146~}
\newcommand{\numGaiaHR}{145~} %one missing BP-RP
\newcommand{\numTwd}{79~}
\newcommand{\numMwd}{47~}
\newcommand{\numTypeWD}{94~}
\newcommand{\numSDSS}{83~}
\newcommand{\numGalex}{92~}
\newcommand{\numIR}{45~}
\newcommand{\numSED}{98~}
\newcommand{\numNoInfo}{49~}
\newcommand{\numHaMH}{97~}

\begin{document} 

%\title{First systematic eROSITA study of cataclysmic variables around the period minimum and a multi-wavelength catalog of period-bounce systems.}
\title{Cataclysmic variables around the period-bounce: \\ An eROSITA-enhanced multi-wavelength catalog}
\titlerunning{Cataclysmic variables around the period-bounce.}

  % \subtitle{I. Overviewing the $\kappa$-mechanism}

   \author{Daniela Muñoz-Giraldo
          \inst{1}
          \and
          Beate Stelzer\inst{1}
          \and
          Axel Schwope\inst{2}
          }

   \institute{Institut f\"ur Astronomie und Astrophysik, Eberhard-Karls Universit\"at T\"ubingen, Sand 1, 72076 T\"ubingen, Germany \\   
   \email{munoz-giraldo@astro.uni-tuebingen.de} 
         \and
          Leibniz-Institut für Astrophysik Potsdam (AIP), An der Sternwarte 16, 14482 Potsdam, Germany\\  
             }

   \date{Received XX; accepted XX}

% \abstract{}{}{}{}{} 
% 5 {} token are mandatory
 
  \abstract
  % context heading (optional)
  % {} leave it empty if necessary  
   {Cataclysmic variables with degenerate donors which have evolved past the period minimum, also known as period-bouncers, are predicted to make up a great portion of the cataclysmic variable population, between 40$\%$ and 70$\%$. However, either due to shortcomings in the models or due to the intrinsic faintness of these strongly evolved systems, only a few have been confidently identified so far.}
  % aims heading (mandatory)
   {We have compiled a multi-wavelength catalog of period-bouncers and cataclysmic variables around the period minimum from the literature in order to provide an in-depth characterization of the elusive subclass of period-bounce CVs that will help in the identification of new candidates.}
  % methods heading (mandatory)
   {In this study we combine published or archival multi-wavelength data with new X-ray observations from the all-sky surveys carried out with the extended ROentgen Survey with an Imaging Telescope Array (eROSITA) onboard the {\it Spektrum-Roentgen-Gamma} spacecraft (SRG). Our catalog comprises 192 cataclysmic variables around the period minimum that were chosen as likely period bounce candidates based on reported short orbital periods and low donor mass. This sample helped us establish specific selection parameters that have been used to compile a “scorecard” which rates a system’s likelihood of being a period-bouncer. 
   %To better characterize this under-represented class we searched for evidence of accretion from their X-ray emission in eROSITA data. 
   }
  % results heading (mandatory)
   {Our "scorecard" correctly assigns high scores to the already confirmed period-bouncers in our literature catalog and it identifies 80 additional strong period-bounce candidates present in the literature that have not been classified as such. We established two selection cuts based on the X-ray-to-optical flux ratio ($-1.21 \leq \rm log(F_x/F_{opt}) \leq 0$) and the typical X-ray luminosity (log(L$_{\rm x,bol})\leq$ 30.4 [erg/s]) observed from the 8 already confirmed period-bouncers with eROSITA data. These X-ray selection cuts led to the categorization of 5 systems as new period-bouncers, increasing their population number to 22 systems.}
  % conclusions heading (optional), leave it empty if necessary 
   {Our multi-wavelength catalog of cataclysmic variables around the period minimum compiled from the literature together with X-ray data from eROSITA resulted in a $\sim30\%$ increase in the population of period-bouncers. Both the catalog and "scorecard" that we constructed will aid in future searches for new period-bounce candidates with the goal of resolving the discrepancy between the predicted high number of period-bouncers and the low number of these systems that have been observed to date.}


   \maketitle
%
%-------------------------------------------------------------------

\section{Introduction}\label{sect:intro}

% CV introduction - definition (Pala Inight paper) 

Cataclysmic variables (CVs) are interacting compact binaries where a white dwarf (WD) accretes matter from a Roche-lobe filling, late-type donor \citep{warner2003}. According to \cite{paczynski1976}, CVs are the result of a common envelope (CE) phase in the evolution of the binary, where the envelope of a Roche-lobe filling, more massive primary star expands enough to engulf the companion. The envelope is then ejected from the binary, leaving a post-common envelope binary composed by the evolved core of the primary (now a WD) and a low mass companion still on the main-sequence. At the end of the CE phase the orbital separation of the binary system has been significantly reduced due to friction in the CE that extracts angular momentum and energy from the system. Once the binary separation is close enough to allow mass transfer onto the WD, the system morphs into a CV.

% early evolution up until period gap - AML from MB - introduction of period gap

In terms of evolution, all CVs follow a track from longer orbital periods towards shorter ones driven by angular momentum loss which causes the orbital separation, and therefore the orbital period, of the system to decrease \citep{knigge2011}. For systems with long orbital periods (P$_{\rm orb}\,>\,$3h) the dominant angular momentum loss mechanism according to the "standard model" is magnetic braking, arising from the stellar wind associated with the secondary's magnetic activity (see e.g \citealt{mestel1968}, \citealt{verbunt1981}). Evolution of the CV continues until an orbital period of around 3h, when the secondary becomes fully convective and, according to the standard model, magnetic braking abruptly stops causing a reduced mass transfer rate in the system that eventually leads to the secondary detaching from its Roche-lobe \citep{knigge2006}. This marks the beginning of the "period gap" (3h $\lesssim$ P$_{\rm orb}\lesssim\,$2h), which contains detached CVs that in the standard model lose angular momentum exclusively from gravitational radiation \citep{spruit1983}. At the lower boundary of the gap, the secondary is filling its Roche-lobe once again allowing for the re-start of mass transfer in the system \citep{howell2001}. 

% short-period evolution after period gap - AML mechanisms from theory (exclusive GR) - does not adjust to observational evidence - Knigge 2006? study finds that you need 2.54GR? to recreate observed evolution - enhanced AML due to additional AML mechanisms still debated lots of "recepies" - talk about CAML and eCAML? Schreiber 2016

According to this paradigm, the system re-emerges from the period gap as a short-period CV with angular momentum loss driven exclusively by gravitational radiation 
%according to the standard model 
\citep{paczynski1976}. However, the evolution track generated by considering gravitational radiation as the only mechanism of angular momentum loss does not comply with observations.
%not fit observational evidence very well. 
\cite{knigge2011} showed that in order to reproduce the observed evolution of short-period CVs, the angular momentum loss mechanism has to be around 2.5 times stronger than pure gravitational radiation. The origin and nature of this enhanced mechanism for angular momentum loss in short-period CVs is a topic of active discussion with a wide variety of "recipes" being suggested (\citealt{politano1996}, \citealt{zorotovic2011}, \citealt{wijnen2015}, \citealt{belloni2020}). Amongst them we highlight the empirical model for consequential angular momentum loss \citep{schreiber2015} which seems to solve several major disagreements between theory and observations, even though the physical mechanism behind the additional angular momentum loss is unclear.  

% short-period CV types - magnetic / non-magnetic (fractions of each?) - containing such groups of CVs as polars, intermediate polars, dwarf novaes (see Inight for how to handle this part/ descriptions)

Both magnetic and non-magnetic systems can be found amongst short-period CVs. Magnetic systems which make up around one third of CVs (\citealt{wickramasinghe2000}, \citealt{pretorius2013}, \citealt{pala2020}) can be further categorized depending on the strength of their magnetic field into polars, with $B \geq 10$MG which suppresses the formation of an accretion disk forcing the accretion flow to follow the magnetic field lines \citep{cropper}, and intermediate polars, with $1 \leq B \leq 10$MG which allows a vestigial disk to form \citep{patterson1994} while accretion remains primarily through the magnetic field lines.

% At this point CVs arrive at period minimum - observational evidence - physical characteristics
% because of new composition of donor system reverses evolution - period bouncers
Short-period CVs, below the period gap, continue to evolve towards even tighter orbits until the system reaches a period minimum. At this point the degenerate donor is out of thermal equilibrium due to its mass-loss timescale becoming much shorter than its thermal timescale, causing the donor to stop shrinking in response to mass-loss \citep{king1988}. The donor, not being able to sustain hydrogen burning, becomes a brown dwarf \citep{howell2001}. This change in internal structure results in the increase of the system's orbital separation and consequently the CV bouncing back to longer orbital periods. The systems that go through this process of "bounce-back" to longer periods are known as period-bouncers \citep{patterson1998}. %paczynski1983

% Period spike - observed with SDSS - indicates a slow evolution through period minimum
Even though the presence of a period minimum has always been a defining characteristic of theoretical models describing CV evolution (\citealt{paczynski1983}, \citealt{howell2001}, \citealt{rappaport1982}, \citealt{kolb1999}), evidence supporting its existence only came with a Sloan Digital Sky Survey (SDSS) study of CVs by \cite{gansicke2009}. A period spike at 80min $\lesssim$ P$_{\rm orb}\lesssim\,$86min was observed providing clear evidence of a "pile-up" of systems that are slowly evolving through the period minimum. 

% population studies of CV - predicted number of PBs - rarely observed (observational bias)
% associated with observational characteristics of PBs - old evolved CV / low luminosity / faint / low mass transfer systems
%Another theorized, 
One of the key predictions of theoretical models describing CV evolution is that period-bouncers are expected to make-up the majority of the CV population. Fractions between 40$\%$ and 70$\%$ of all CVs should have evolved beyond the period minimum, depending heavily on the formation and evolution model used as well as the assumptions made about the systems parameters (see e.g. \citealt{kolb1993}, \citealt{goliasch}, \citealt{belloni2020}). However, from a volume-limited sample study of CVs by \cite{pala2020}, within 150pc the observed fraction of period-bouncers is only between 7$\%$ and 14$\%$, and to date there are less than 20 confirmed period-bouncers (see e.g. \citealt{patterson2005}, \citealt{mcallister2017}, \citealt{neustroev2017}, \citealt{pala2018}, \citealt{schwope2021}, \citealt{amantayeva2021}, \citealt{kawka2021}, \citealt{munoz2023}).
This under-representation may be due to selection biases against period-bouncers, as their defining characteristics of being old and faint CVs with low luminosity and mass transfer make their detection challenging \citep{patterson2011}.  

% recent association between WZ Sge type objects an PBs - likely source for finding new ones
% most WZ Sge come from photometric studies of optical light curves - observation of superoutbursts (rare) - kato has compiled a large number of Porb and massDonor from this studies - objects remain largely unstudied in other wavelenghts - makes classification difficult
Several recent studies suggest dwarf novae as the most likely source for the missing population of period-bouncers (see e.g. \citealt{uemura2010}, \citealt{kimura2018}). 
These non-magnetic CVs are characterized by the presence of an accretion disk around the WD which can be observationally identified due to the quasi-periodic changes in brightness known as ``outbursts" \citep{meyer1984}.
%
% Very interesting group amongst this is SU UMa which are DN ... populate primarly the short-period regime with a subset of them WZ Sge having the shortest periods of all DN (observational numbers?)
Dwarf novae can be further classified into several sub-classes depending on the behavior of the outbursts. SU UMa systems are characterized for not only exhibiting outbursts, but also superoutbursts which can last up to several weeks \citep{vogt1980}. An interesting, and useful, property of these superoutbursts is that they are accompanied with the presence of "superhumps" which originate from donor induced tidal dissipation of an eccentric accretion disk in a 3:1 orbital resonance (\citealt{whitehurst1988}, \citealt{osaki1989}). The period of these "superhumps", which is expected to be a few percent longer than the orbital period \citep{patterson2005b}, has been established to scale with the mass ratio of the system, making it an ideal tool to estimate the masses of the individual components in the CV (\citealt{patterson2005b}, \citealt{kato2013}). 

The vast majority of SU UMa are located below the period gap, with donors that are usually M-type or later, making it common to describe the overall short-period non-magnetic CV population as SU UMa-type objects (\citealt{kato2009} and further papers in this series).
When an SU UMa has been established through observations to have extremely long outbursts recurrence times (in the order of years to decades), a very low mass transfer rate (in the order of $10^{-11} \rm M_{\odot}/yr$) and a low mass donor, this CV can be further categorized as a WZ Sge system (\citealt{patterson2002}, \citealt{kato2015}). WZ Sge-type CVs have usually the shortest orbital periods of the overall non-magnetic CV population and very low mass donors which may be brown dwarfs \citep{kato2022}. 
In fact, the majority of systems found in the "period spike" are classified as WZ Sge-type CVs \citep{gansicke2009}, which further supports this theory, establishing the population of already identified WZ Sge-type objects as a good source to find new period-bounce candidates. However it is important to consider that these objects are detected primarly through photometric studies of optical superoutburst light curves, meaning that they are not easily observed in quiescence due to their faintness
%\textcolor{orange}{\bf BS-comment: What do you mean? Too faint?} 
and, therefore, have remained largely undetected in other wavelengths, making a straightforward classification as a period-bouncer challenging.

%On the other hand, non-magnetic systems account for the majority of the CV population \citep{pretorius2012}, and included the sub-type of dwarf novae. These non-magnetic CVs are characterized by the presence of an accretion disk around the WD which can be observationally identified due to the quasi-periodic changes in brightness known as outbursts \citep{meyer1984}.

% other source for period-bounce candidates are spectroscopic studies - Pala and Inight study - characteristics by both
% introduces WD appereance of PBs that comes from their already discussed characteristics - dominant WD makes difficult to distinguish detached systems from accreting binaries - Xrays as a key to fix this (proof of accretion)
Spectroscopic studies using a variety of instruments including the {\it Hubble Space Telescope} (HST) and SDSS (see e.g. \citealt{pala2022}, \citealt{inight2023a}) are also an important source for period-bounce candidates. Spectra in the ultra-violet (UV) and optical bands, in the case of period-bouncers are strongly dominated by the WD, often with no contribution from the late-type donor at all. This introduces new possibilities and challenges in the search for period-bouncers as potential candidates could be found in single WD catalogs \citep{inight2023a} in great numbers, but it makes it necessary to distinguish between these systems. Period-bouncers could also be incorrectly categorized as detached binaries, specially considering the low mass transfer rates that they exhibit (\citealt{inight2021}, \citealt{schreiber2023}). V379 Vir is a confirmed period-bouncer that was initially assumed to be a detached binary \citep{schmidt2005}, but it was later proved to be accreting through the detection of orbital modulation from a deep X-ray observations using {\it XMM-Newton} \citep{stelzer2017}. 

% Specific case of SDSS 1212 - believed to be a detached binary (look up reference) - Xray detection by Beate - together with already PB like characteristics that were observed leads to a confident PB classification
% XMM as a tool to do this - advantages: data quantity etc (examples SDSS 1514/ SDSS 1212 / Axel paper) - disadvantage: time consuming / not all candidates will be accepted for observation
Considering that coronal X-ray emission is not expected from the very late-type donors of period-bouncers (\citealt{audard2007}, \citealt{deLuca2020}), detection of X-ray emission is a key diagnostic of ongoing mass accretion in these systems, which unequivocally distinguishes period-bouncers from isolated WDs and detached binaries, and hence is the most promising path for the identification of new candidates. Even though the high sensitivity of X-ray instruments, like {\it XMM-Newton}, is useful when identifying accreting period-bounce systems, the use of all-sky surveys might be more pertinent for the large-scale search of period-bouncers that is required in order to bring the number of detected systems up to the expected values.

% introduces the need for a all sky-survey like ROSAT  - give general specifications
% However not deep enough for faint targets like PBs
% Introduce eROSITA (Axel paper) - covers half the sky / good resolution - opportunity to detect a large number of candidates providing proof of accretion and contributing to the studies of mass accretion rate and luminosity of PBs 
With the launch of the extended ROentgen Survey with an Imaging Telescope Array \citep[eROSITA;][]{predehl2021} onboard the {\it Spektrum-Roentgen-Gamma} mission \citep[SRG;][]{sunyaev2021} we gained access to large statistical samples of X-ray sources. 
The first eROSITA detections of confirmed period-bouncers \citep{munoz2023} proved the capabilities of this instrument in the study and classification of such faint sources, suggesting it as a reliable tool for the identification of new period-bounce candidates. The X-ray study of period-bouncers with eROSITA will help us establishing the class properties that will, in the future, aid the search for new candidates with the aim of resolving the discrepancy between observed and predicted fraction of period-bouncers.

% aim / motivation / goals
The main focus of this study is to produce a catalog of CVs around the period minimum, that is potential period-bouncers, that were already discussed in the literature, in order to characterize in detail the elusive subclass of period-bounce CVs. To this end, we constructed a literature catalog including both confirmed period-bouncers and candidates. The inclusion of confirmed period-bouncers allows us to use their reported parameters to rate how likely the other candidates are of being a period-bouncer. This catalog will also provide a reliable observational data set that could be used for future theoretical modelling of advanced CV evolution.

% summary of individual sections
Our literature catalog of period-bounce candidates is introduced in more detail in Sect.~\ref{sect:catalog}. We provide specific information in the selection of candidates, followed by an overview of the parameter values used from the literature and comments on the evolution of short-period CVs that can be derived from this information. We finish Sect.~\ref{sect:catalog} with a discussion on the quality of the distance values used in the catalog. The process to obtain the photometry of the systems in the catalog is explained in Sect.~\ref{sect:photometry}, In Sect.~\ref{sect:scorecard} we introduce and explain in detail the scorecard we constructed in order to rate the likelihood a system has of being a period-bouncer. In Sect.~\ref{sect:erosita} we discuss the eROSITA X-ray detections of period-bounce candidates from our catalog and we establish their X-ray characteristics. We give our conclusions in Sect.~\ref{sect:conclusions}.

%--------------------------------------------------------------------
\section{Literature catalog of period-bounce candidates}\label{sect:catalog}

We have compiled a catalog of systems already reported in the literature which are characterized by being short-period CVs that are close to or may already have gone through the period minimum.
%for CVs. 
This collection of CVs around the period-bounce constitutes an increase in the number of systems of about an order of magnitude compared to earlier studies of period-bouncers (\citealt{littlefair2003}, \citealt{knigge2006}, \citealt{gansicke2009}, \citealt{patterson2011}, \citealt{inight2023a}), and provides
a comprehensive overview of what is currently known of CVs in the short-period regime.

\subsection{Candidate selection}\label{subsect:candidateSelection}

We have manually gone through different databases and individual papers in order to compile a "unified" catalog of potential members of the widely elusive subclass of CVs known as period-bouncers. To this end, we enforced that the selected objects had both a very low donor mass and an orbital period placing them around the period minimum or in the post-bounce area. Additionally, we kept systems that were suggested as good period-bounce candidates even if they had no reported parameters or their literature parameters did not match our above criteria on donor mass and system orbital period. Examples of this include systems resembling period-bouncers when analysing their spectrum \citep{inight2023a}, optical light curve \citep{kimura2018}, WD temperature \citep{pala2022}, among others, even if their orbital period or donor mass could not be determined probably due to sparse or poor quality data.
%data, often of bad quality. 

The data sources from which we got most of the systems are in order of date of publication: \cite{littlefair2008},  \cite{patterson2011}, \cite{otulakowska2016}, \cite{kimura2018}, \cite{longstaff2019}, \cite{kato2022} and \cite{inight2023a}, from now on referred to as "Literature Catalogs". Additionally, we also considered systems that were studied individually in dedicated articles (see e.g. \citealt{savoury2012}, \citealt{beuermann2021}, \citealt{kawka2021}, \citealt{wild2022} and \citealt{liu2023}). This resulted in a literature catalog of \numTotal systems, by far the largest data base of period-bounce candidates studied to date. 
%containing a growing number of systems, \numTotal to date.
%\textcolor{orange}{\bf BS-comment: What is growing here? This is your final catalog !}\textcolor{red}{I wanted to comment that given the conditions of Porb and Mdonor that we set the catalog was expanded several times and could technically still be expanded on.}

The vast majority of the systems in our catalog, 133 out of \numTotal candidates, come from photometric studies of dwarf novae (both SU~UMa and WZ~Sge-type objects) specifically aimed at the study of superhumps in their superoutbursts (\citealt{otulakowska2016}, \citealt{kimura2018}, \citealt{kato2022}). This type of photometric studies are specially relevant for us because they provide a superhump period for a large number of systems that can be used to obtain the mass ratio of the system (see \citealt{kato2013} for a comprehensive review of this method), and therefore yields the mass of the donor for a known WD mass.

\begin{figure}
   \centering
   %trim={<left> <lower> <right> <upper>}
    \includegraphics[width=\columnwidth]{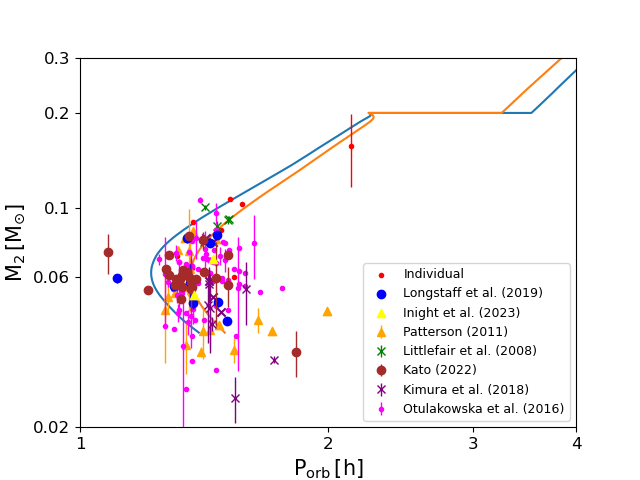}
    \caption{Donor mass as a function of orbital period for the objects in our literature catalog of period-bounce candidates. The color and shape corresponds to the catalogs from which we took the system (and that are listed in the legend). Two CV evolution tracks from \cite{knigge2011} are shown as reference: the standard track considering angular momentum loss only due to gravitational radiation (blue) and a revised track that considers enhanced angular momentum loss (orange).}
    %: green x for \cite{littlefair2008}, orange triangle for \cite{patterson2011}, small magenta dot for \cite{otulakowska2016}, purple x for \cite{kimura2018}, large blue dot for \cite{longstaff2019}, large brown dot for \cite{kato2022}, yellow triangle for \cite{inight2023} and small red dot for systems coming from individual references.}
    \label{PorbMdonor}
\end{figure}

The location of the \numPorbMdonor systems 
that have both orbital period and donor mass reported in the literature is presented in Fig.~\ref{PorbMdonor} with respect to the period-bounce area predicted from different tracks for the evolution of CVs by \cite{knigge2011}. The error bars presented in Fig.~\ref{PorbMdonor} represent the range of donor masses measured for the systems if more than one value was found in the literature. All of the systems except for one, V1258 Cen, populate the area around the period minimum with a sizeable number of them located in the post-bounce area. V1258 Cen was proposed as a period-bouncer by \cite{pala2022} due to its very low WD temperature, a key characteristic of this type of CV. However, detailed studies of V1258 Cen rule out this system as a period-bouncer, considering that it has an orbital period of 128min \citep{mcallister2019} and a donor mass of 0.198 $\pm$ 0.029\,M$_{\odot}$ \citep{savoury2012} which are too large to belong to a system past the period minimum %\textcolor{orange}{\bf BS-comment: What about Porb of V1258Cen?}. 
We decided to keep this system in the catalog in order to compare its characteristics to other systems that are more likely to be period-bouncers. In particular, V1258 Cen may serve as a check to the scorecard that we define in Sect.~\ref{sect:scorecard}.

%                                                One column figure
%----------------------------------------------------------------- 
%-----------------------------------------------------------------

%Most importantly, 
Our literature catalog contains \numPB "bona fide" period-bouncers. These systems, shown in Table~\ref{confirmedPB}, have been classified in the literature as such because they exhibit several key characteristics of period-bouncers in addition to a spectroscopic or photometric confirmation of a late-type donor. These confirmed period-bouncers will be used as a check to the scorecard, where col.~4 of Table~\ref{confirmedPB} anticipates the final score for each of them according to the analysis we present in Sect.~\ref{sect:scorecard}.  In our scoring system the value $100$\,\% is given to systems that have achieved the maximum possible score. Only two of the confirmed period-bouncers reach a 100$\%$ score, with a minimum score of 64$\%$ which defines the lower boundary that we apply to the full catalog in Sect.~\ref{sect:scorecard}.

\begin{table}
    \begin{threeparttable}
    \centering
    \caption{
    %\textcolor{red}{(1) Are these 17 systems identified in the plots? (2) "BD" is not a spectral type.} MISSING
    Confirmed period-bouncers in the literature catalog. We include the reported spectral type of the donor with the method used for its confirmation and the corresponding reference. The last column refers to the final score of each system defined in Sect.~\ref{sect:scorecard} to estimate their likelihood of being a period-bouncer.}
    \label{confirmedPB}
    \begin{tabular}{cccc} 
    \hline
    \noalign{\smallskip}
    Period-bouncer & Donor SpT & Reference & Score [$\%$] \\
    \noalign{\smallskip}
    \hline
    \noalign{\smallskip}
    V379 Vir & L8 (S) & (1) & 100 \\
    SDSS 1514 & L3 (P) & (2) & 69 \\
    SDSS 1250 & M8 (P) & (2) & 67 \\
    SDSS 1057 & L5 (P) & (3) & 86 \\
    SDSS 1433 & L1 (S) & (4) & 72 \\
    WZ Sge & L2 (S)& (5) & 94  \\
    SDSS 1035 & L0 (P) & (6) & 86 \\
    SMSS 1606 & L2 (P) & (7) & 73 \\
    QZ Lib & BD (S) & (8) & 100 \\
    GD 552 & BD (S) & (9) & 80 \\
    MT Com & Very late (P) & (10) & 77 \\
    V455 And & L2 (P) & (11) & 89 \\
    V406 Vir & L3 (S) & (12) & 64 \\
    BW Scl & T (S) & (13) & 83 \\
    EZ Lyn & L2 (P) & (14) & 97 \\
    SSS 1222 & L0 (P) & (15) & 80 \\
    V1108 Her & L1 (S) & (16) & 87\\
    \hline
    \end{tabular}
    \begin{tablenotes}
        \small
        \item Confirmation of brown dwarf (BD) or late-type donor through spectroscopy (S) or photometry (P).
        \item {\it References.} (1) \cite{farihi2008}, (2) \cite{breedt}, (3) \cite{mcallister2017}, (4) \cite{santisteban2016}, (5) \cite{harrison2015}, (6) \cite{schwope2021}, (7) \cite{kawka2021}, (8) \cite{pala2018}, (9) \cite{unda2008}, (10) \cite{patterson2005}, (11) \cite{araujo2005}, (12) \cite{pala2019}, (13) \cite{neustroev2022}, (14) \cite{amantayeva2021}, (15) \cite{neustroev2017},(16) \cite{ishioka2007}.
    \end{tablenotes}
  \end{threeparttable}
\end{table} 

%Even though all the objects in this sub-sample are named in the literature as period-bounce candidates, only 14 of them have been definitely categorized as period-bouncers: V379\,Vir (L8 donor; \citealt{farihi2008}, \citealt{stelzer2017}), SDSS\,1433 (L1 donor; \citealt{santisteban2016}), WZ\,Sge (L2 donor; \citealt{steeghs2007}), QZ\,Lib (brown dwarf donor; \citealt{pala2018}), GD\,552 (brown dwarf donor; \citealt{unda2008}), SDSS\,1238 (L3 donor; \citealt{pala2019}), BW\,Scl (T donor; \citealt{neustroev2022}), V1108\,Her (L1 donor; \citealt{ishioka2007}, \citealt{pavlenko2011}), SDSS\,1514 (L3 donor; \citealt{breedt}, \citealt{munoz2023}), SDSS\,1250 (M8 donor; \citealt{breedt}), SDSS\,1057 (L5 donor; \citealt{mcallister2017}), SMSS\,1606 (L2 donor; \citealt{kawka2021}), EZ\,Lyn (L2 donor; \citealt{amantayeva2021}), SSS\,1222 (\citealt{neustroev2017}). 

\subsection{Literature values}\label{subsect:literatureValues}

We aimed at compiling a catalog that has as much information on the period-bounce candidates as possible, including literature values on the following parameters: name, {\it Gaia}-DR3 ID, coordinates (J2000), orbital period ($P_{\rm orb}$), temperature of the WD ($T_{\rm eff}$), mass of the WD ($M_{\rm WD}$), WD magnetism type, mass of the donor ($M_{\rm donor}$), spectral type of the donor (SpT$_{\rm donor}$), and the {\it Gaia}-DR3 distances given by \cite{bailer2021}. These parameters were drawn for the most part from the Literature Catalogs listed in Sect.~\ref{subsect:candidateSelection}.
%, if they were available. 
In the case where no information was given on some parameters, or an updated value existed, we used additional references to supplement the catalog. A shortened version of our literature catalog is shown in Table~\ref{PBtable} with a brief description of the columns in Table~\ref{PBtable_col}. In the following we specify how we chose them from the literature. 

{\it Orbital period, available for \numPorb systems.} The majority of orbital period values were derived through light curve analysis from optical surveys (see e.g. \citealt{otulakowska2016}, \citealt{kato2022}), with a few of them coming from other methods like radial velocities from H$\alpha$ emission (see e.g. \citealt{breedt}, \citealt{pala2018}) or X-ray light curves (see e.g. \citealt{stelzer2017}, \citealt{munoz2023}). In the specific instances where more than one source had a reported orbital period for a given system, the values were in all cases consistent with each other. In these cases, we chose to keep the value from the most recent derivation. 

{\it Donor mass, available for \numPorbMdonor systems.} Considering that this parameter has the highest degree of uncertainty, as it is rarely measured directly, when possible we used two values for each object, often determined from different methods. This allowed us to obtain a range for the expected mass of the secondary, illustrated by the error bars in Fig.~\ref{PorbMdonor}. The majority of the values come from using the WD mass together with the mass ratio ($q = M_{\rm donor}/M_{\rm WD}$) determined by either the superhump method \citep{kato2013} or the eclipse modeling method \citep{savoury2011}. If no value for the WD mass was available, we used 0.8M$_\odot$ which was determined by \cite{pala2022} to be the mean mass of WDs in CVs.

{\it Spectral type of the donor, available for \numSpT systems.} An accurate determination of the SpT is particularly relevant for the study of period-bouncers as the spectral type of the donor -- together with its mass -- is the parameter that is both the most important and the most difficult to determine precisely in a system with faint, very-low mass donor. We used the classifications that had preferably a spectroscopic (see e.g. \citealt{farihi2008}, \citealt{santisteban2016}) or, if not, a photometric confirmation (see e.g \citealt{amantayeva2021}, \citealt{kawka2021}). 

{\it WD temperature, available for \numTwd systems.} The majority of the values come from UV spectroscopy \citep{pala2022} and optical photometry \citep{gentile2021}. When no value was available from either of the methods we considered temperatures derived from eclipse modeling or evolutionary status (see e.g. \citealt{mcallister2019}, \citealt{savoury2011}). 

{\it WD mass, available for \numMwd systems.} The majority of the values come from eclipse modeling (see e.g. \citealt{savoury2011}; \citealt{mcallister2019}) and UV spectroscopy \citep{pala2022}. When neither value was available we considered masses derived from gravitational redshift \citep{neustroev2022} or optical spectroscopy \citep{munoz2023}. 

{\it WD magnetism type, available for \numTypeWD systems.} From broad sample studies (see e.g. \citealt{patterson2011}, \citealt{belloni2020}, \citealt{pala2022}) which commonly focus on non-magnetic CVs as they are better understood and expected to follow standard evolution tracks, we categorize the systems present in these studies as being non-magnetic. For magnetic CVs we used individual references that in most cases identified the magnetic nature of the WD from the optical spectra (see e.g. \citealt{breedt}, \citealt{kawka2021}).
%\textcolor{red}{If you say that you use indiviual references you must list them all.} MISSING

\subsection{Comments on short-period CVs}\label{subsect:CVevolution} 

Considering that our catalog of short-period CVs has a large population of systems with information about key parameters in CV evolution, we use this knowledge to comment on the observational characteristics of this type of systems.

\subsubsection{Evolution tracks}\label{subsubsect:evolutionTracks} 

The evolution of CVs, especially in the short-period regime, is still heavy debated, with theoretical models often clashing with results from observations. This has led to the development of several empirical (or semi-empirical) evolution tracks.
%in order to adjust better to observations.  
\cite{knigge2011} proposed a semi-empirical donor-based CV evolution track that in the short-period regime relies on a mechanism for angular momentum loss that is around 2.5 times stronger than pure gravitational radiation. The difference this change introduces can be easily observed in Fig.~\ref{PorbMdonor}, where the revised model (orange line) is characterized by a period-bounce area associated with longer orbital periods when compared to the standard theoretical model (blue line). 
From the \numPorbMdonor systems in our catalog of period-bounce candidates that have a literature value for both the orbital period and the donor mass (shown in Fig.~\ref{PorbMdonor}), we notice that
%, from observations, the system parameters of 
short-period CVs seem to be located preferentially around the revised \cite{knigge2011} %revised 
evolution track. Only a few of the period-bounce candidates reach the short orbital periods associated with the period minimum of the standard evolution track, showing that for a given donor mass this track substantially underestimates the orbital period of the system. 
%\textcolor{orange}{\bf BS-comment: Is this surprising? Wasn't the Knigge revised track designed to fit the observed period minimum?} \textcolor{red}{\bf It is not surprising but it does serve as a check for this evolution track considering that Knigge sample is composed mainly by CVs with longer periods with very few systems at the period bounce area, as seen in their Figure 9. He even talks about "poorly constrained period-bounce regime" due to the lack of physical models that accurately describe the evolved donors in these systems.} 
However, it is important to note the considerable scatter in the distribution of the systems at or after the period minimum suggesting that the unique track of CV evolution might diverge around the bounce-back area. This scatter has been suggested to be related with the substantial range of masses that the WD may have \citep{howell2001} as the WD mass in CVs has been proven to not be constrained by the orbital period \citep{pala2022}. 

\subsubsection{WD magnetism}\label{subsubsect:WDmagnetism} 

\begin{figure}
    \centering
    %trim={<left> <lower> <right> <upper>}
    \includegraphics[width=\columnwidth]{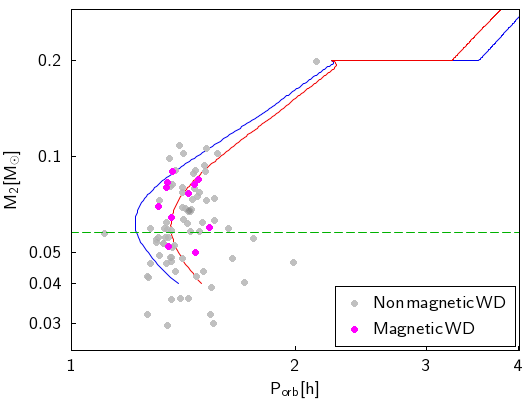}
    \caption{Evolution for magnetic and non-magnetic CVs for the \numTypeWD systems in our literature catalog with information about the magnetism of the WD. Point of period reversal (M$_{\rm donor} \approx 0.058 \rm M_{\odot}$) is shown by the green dashed line. The two \cite{knigge2011} CV evolution tracks from Fig.~\ref{PorbMdonor} are shown as reference.}
    \label{WDmagnetism}
\end{figure}

Magnetism of the WD has been considered as a factor that may affect the evolution of CVs, leading to different evolution tracks depending on whether the WD is strongly magnetic or not \citep{gansicke2009}. In magnetic CVs %the effects of 
magnetic braking is expected to be reduced as the wind from the donor may be trapped within the magnetosphere of the WD, causing a difference in the rate of angular momentum loss between magnetic and non-magnetic CVs \citep{webbink2002}. However, this difference is expected to be significantly reduced for short-period CVs as, in the standard model, gravitational radiation is believed to be the primary mechanism of angular momentum loss in the system with only a weak contribution by magnetic braking \citep{schreiber2021}. The distribution of the \numTypeWD systems in our catalog with information about the magnetism of the WD support this scenario, considering that we do not observe an evolution track in Fig.~\ref{WDmagnetism} being preferentially populated by either magnetic or non-magnetic CVs. 

%----------------------------------------------------------------- 
%-----------------------------------------------------------------

What we do observe from Fig.~\ref{WDmagnetism} is a clear difference between the pre- and post-bounce distribution of the magnetic and non-magnetic CVs. Only 2 out of the 11 magnetic systems are clearly past the point of period reversal (M$_{\rm donor} \approx 0.058 \rm M_{\odot}$; \citealt{knigge2011}) corresponding to $\sim 20\%$ of this population. On the other hand 34 out of 82 non-magnetic systems are located in the period-bouncer area accounting for $\sim 40\%$ of this population. This implies that non-magnetic CVs are twice as likely to have evolved past the period minimum than magnetic CVs which might be taken as a hint for longer evolutionary times for magnetic CVs.

The lack of differentiation between evolution tracks of magnetic and non-magnetic CVs, as well as the overrepresentation of magnetic CVs in the pre-bounce area, could also be explained by the low number of observed short-period magnetic CVs, which leads to an incomplete representation of their evolution track.
%\textcolor{orange}{\bf BS-comment: What do you want to say?} \textcolor{red}{\bf That the two characteristics could also be due to a very incomplete population of magnetic CVs and would "disappear" once we have the complete population} MISSING
Magnetic CVs in our catalog account for 10$\%$ of the population which is considerably less than the expected $\sim$30$\%$ (\citealt{wickramasinghe2000}, \citealt{pretorius2013}, \citealt{pala2020}) obtained in previous CV population studies. Rather than showing the true picture of magnetic and non-magnetic short-period CVs, this result evidences a bias of our catalog towards non-magnetic systems.
As already discussed in Sect~\ref{subsect:candidateSelection} the majority of our candidates are dwarf novae, a type of non-magnetic CV, selected from large photometric studies that provide both orbital period and mass of the systems. Such databases are not as readily available for magnetic systems making their inclusion into our catalog more challenging as we had to rely on individual papers that often did not include all the parameters required. This results in and underrepresentation of magnetic CVs in our catalog of period-bounce candidates.

\subsubsection{Period spike and period minimum}\label{subsubsect:minimum} 

The number density of CVs at a given period is inversely proportional to the rate at which the orbital period evolves \citep{littlefair2008}, making a pile-up or "spike" of systems at the period minimum an expected characteristic of CV evolution. This was observed for the first time by \cite{gansicke2009} using 137 CVs from SDSS. The period "spike" was established to be located in the orbital period range of 80min to 86min corresponding to a significant accumulation of CVs with P$_{spike}$ = 82.4$\pm$0.7 min. 

\begin{figure}
    \centering
    %trim={<left> <lower> <right> <upper>}
    \includegraphics[width=\columnwidth]{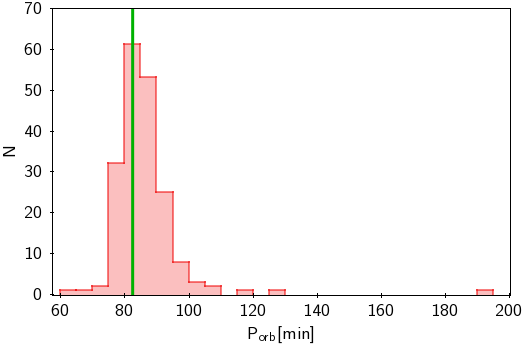}
    \caption{Distribution of the orbital period for \numPorb systems in our literature catalog with values from the literature. The vertical green line is an estimate of P$_{spike}$ using a Gaussian fit to the distribution of systems with orbital periods between 77min and 87min.} 
    \label{Hist}
\end{figure}

Fig.~\ref{Hist} shows the orbital period distribution for \numPorb period-bounce candidates in our catalog with a literature value. The period "spike" is observed within the expected range of 80min to 86min. We estimated P$_{spike}$ = 82.4$\pm$2.7 min (see green line in Fig.~\ref{Hist}) following the description of \cite{mcallister2019} who used a Gaussian fit to the distribution of systems with orbital periods between 77min and 87min to obtain P$_{spike}$ = 82.7$\pm$0.4 min. Both values are consistent with each other as well as 
%consistent 
with the \cite{gansicke2009} value.  Our estimation for P$_{spike}$ is also consistent with the 81.8$\pm$0.9 min period minimum predicted by \cite{knigge2011} in their revised donor track. 

%----------------------------------------------------------------- 
%-----------------------------------------------------------------

\subsection{Distance estimates and limits}\label{subsect:distance} 

To determine the optical properties, and especially the distances, to the systems we performed a match between our literature catalog and the {\it Gaia}-DR3 catalog\footnote{\url{https://cosmos.esa.int/web/gaia-users/archive}}. Hereby we checked that the resulting match coincided with the {\it Gaia}-DR3 ID available in SIMBAD\footnote{SIMBAD is the Set of Identifications, Measurements and Bibliography for Astronomical Data (\url{http://simbad.cds.unistra.fr/simbad/)}}. Of the \numTotal systems in our catalog \numGaia have available {\it Gaia}-DR3 data. \numGaiaBJ of our candidates have {\it Gaia} detections at different epochs meaning that they have an available parallax \citep{gaiadr3} that can be used to estimate the distance to the systems.

\begin{figure}
    \centering
    %trim={<left> <lower> <right> <upper>}
    \includegraphics[width=\columnwidth]{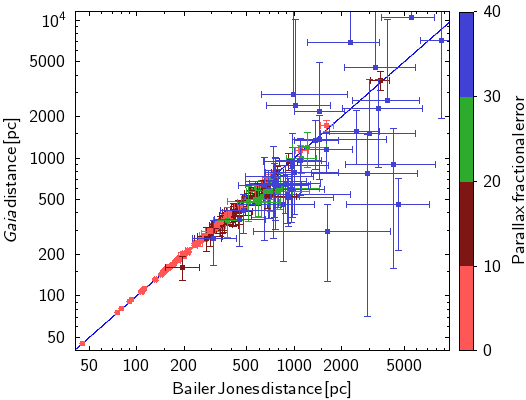}
    \caption{Comparison between the distance obtained by \cite{bailer2021} and by using the inverse of the parallax.}
    \label{distaceBJplx}
\end{figure}

However, using parallaxes to estimate distances is not a trivial process considering that simply using the inverse of the parallax introduces biases in the distance estimate which affect especially systems with large fractional errors \citep{pala2022}. As can be seen in Fig.~\ref{distaceBJplx}, the more distant among the systems from our catalog present large fractional errors, which are greater than $20\%$ for $49$ of our period-bounce candidates 
%\textcolor{orange}{\bf BS-comment: Is that the parameter you mean? {\sc parallax\_over\_error}}\textcolor{red}{No, I mean $(Plx_{Error}/Plx)*100$}. 
This is not an unexpected result considering that these systems have on average $G\sim 20.1$mag, making their intrinsic faintness, which is characteristic of period-bouncers, the most likely cause for the large uncertainties of their parallaxes (see Fig.~7 from \citealt{lindegren2021}). 
%\textcolor{orange}{\bf BS-comment: I'm not sure. How large/small are the errors for brighter stars at similar large distance?} \textcolor{red}{It can be seen even with our sample. The brown system at 4000pc has 16.2 mag, while the pink systems at 1000pc/1500pc have 16.2mag and 17.5mag. All the green and blue systems with large distances have 19.3mag or larger.}

In order to avoid the introduction of bias into our calculation, which would affect $49$ of our candidates with a {\it Gaia} parallax value, we chose to use the distance estimates obtained by \cite{bailer2021}, specifically the geometric distance which takes a probabilistic approach that uses a prior constructed from a three-dimensional model of our Galaxy to estimate the distances (see \citealt{bailer2015} for more information on this method). Fig.~\ref{distaceBJplx} illustrates the difference between both methods used to estimate the distance, showing a significant disagreement for systems at large distance where the distance from the parallax is underestimated with respect to the distances obtained by \cite{bailer2021}.

%----------------------------------------------------------------- 
%-----------------------------------------------------------------

Having an accurate value for the distance is especially relevant for studies like ours considering that many system parameters depend on distance. An example of this is the magnetic CV EF Eri, which was considered a good period-bounce candidate because at an estimated distance of 111pc the photometry suggested a brown dwarf donor in the system \citep{schwope2010}. However, once the distance was revised to 160pc \citep{bailer2021}, this argument was not valid anymore disqualifying EF Eri from being a good period-bounce candidate. {As EF Eri is included in our literature catalog, we use it similarly to V1258 Cen as a check to our score card in Sect.~\ref{sect:scorecard}.} 

\section{Photometry}\label{sect:photometry}

One possible discriminator between period-bouncers and other CVs regards the IR excess in the composite spectral energy distribution (SED) of the binary system. We include the IR excess as one of the parameters that define our period-bouncer scorecard (Sect.~\ref{sect:scorecard}, item\,10). Therefore, we first had to construct the SEDs for the systems from our literature catalog of period-bounce candidates. We used the {\it Tool for OPerations on Catalogues And Tables}  (TOPCAT\footnote{\url{https://www.star.bristol.ac.uk/mbt/topcat/}}; \citealt{taylor2005}) in order to handle our extensive catalog. TOPCAT is a JAVA-based interactive graphical viewer and editor for tabular data that facilities the analysis and manipulation of source catalogues, specially the cross-match between catalogs.

We performed a cross-match in TOPCAT between our catalog and different photometric surveys covering a wide range of wavelengths from ultraviolet (UV) to infrared (IR). When performing the match with the algorithm ‘sky’ we used the right ascension (RA) and declination (DEC) of each system in coordinates J2000, allowing for a match radius of 30".

We used the following surveys with the latest catalog version available in TOPCAT: Galaxy Evolution Explorer (GALEX) general release GR6+7 \citep{bianchi2017}; Sloan Digital Sky Survey (SDSS) data release DR16 \citep{ahumada2022}; UKIRT Infrared Deep Sky Survey (UKIDSS) Large Area Survey data release DR9 \citep{lawrence2007}; Two Micron All Sky Survey (2MASS) \citep{skrutskie2006}; VISTA Hemisphere Survey (VHS) data release DR5 \citep{mcmahon2021}; Wide-field Infrared Survey Explorer mission (WISE) \citep{cutri2021}. After obtaining the best matches for each system using a 30" radius, we carried out a visual check to make sure that there were no other {\it Gaia} sources in the vicinity. We discarded those of the multi-wavelength matches that most likely do not belong to our candidate, as they are associated with another {\it Gaia} source. This allowed us to obtain an optimal radius with which at least 94$\%$ of the correct matches were retrieved and at maximum one incorrect match:
%\textcolor{orange}{\bf BS-comment: How do you know about the 94\% of correct matches?} \textcolor{red}{I calculated it from the number of correct matches within 30arcsec and within the chosen period}
4" for GALEX (103 correct associations), 4" for SDSS (82 correct associations), 1" for UKIDSS (27 correct associations), 2" for 2MASS (32 correct associations), 3" for VHS (46 correct associations) and 4" for WISE (94 correct associations). The multi-wavelength photometry we collected this way can be found in the online version of our catalog, with the column descriptions listed in Table~\ref{PBtable_col}. 

Using the available photometry for each system we constructed their individual spectral energy distributions (SEDs), ideally covering wavelengths from UV to IR. These SEDs are used in Sect.~\ref{sect:scorecard}, to search for an IR excess and to evaluate if it could be attributed to the presence of a very late-type donor. 

%\section{Scorecard}\label{sect:scorecard}
\section{Period-bouncer likelihood}\label{sect:scorecard}

We used the parameters of the candidates reported in the literature catalog to rate how likely a system is to be a period-bouncer. In this evaluation we considered 10 parameters: spectral type of the donor, donor mass, orbital period, WD temperature, photometric colors (UV, optical and IR), optical variability, and IR excess. We compiled them into a scorecard in which we assigned different weights to the 10 parameters depending on how relevant we judged them for confirming a candidate as a period-bouncer. A summary of the parameters and their respective scores is presented in Table~\ref{scorecardParameters} and is described in detail below.

\begin{table}[h!]
    \centering
    \caption{Scores assigned for the different parameters. See text for full descriptions of each parameter.}  
    \label{scorecardParameters}
    \begin{tabular}{rl | c } 
    \hline
    \noalign{\smallskip}
    & & \textbf{Points}\\
    \noalign{\smallskip}
    \hline
    \noalign{\smallskip}
    \multicolumn{2}{c|}{\textbf{SPECTRAL TYPE}}  & \\
    \ldelim\{{3}{18.2mm}[Spectroscopic] & T/L donor & 9\\
    & M5 or later donor & 5 \\
    & Very late donor & 2 \\
    \ldelim\{{3}{25mm}[Non-spectroscopic] & T/L donor& 7 \\
    & M5 or later donor& 3 \\
    & Very late donor& 1 \\
    \multicolumn{2}{c|}{Other}& 0 \\
    \noalign{\smallskip}
    \hline
    \noalign{\smallskip}
    \multicolumn{2}{c|}{\textbf{DONOR MASS}}& \\
    \multicolumn{2}{c|}{M$_{\rm donor} \leq 0.058 \rm {M_{\odot}}$} & 3 \\
    \multicolumn{2}{c|}{$0.058\rm {M_{\odot}} <$ M$_{\rm donor} \leq 0.07 \rm {M_{\odot}}$} & 1 \\
    \multicolumn{2}{c|}{M$_{\rm donor} > 0.07 \rm M_{\odot}$} & 0 \\
    \noalign{\smallskip}
    \hline
    \noalign{\smallskip}
    \multicolumn{2}{c|}{\textbf{ORBITAL PERIOD}} &\\
    M$_{\rm donor} \leq 0.058 \rm M_{\odot}$ & For any P$_{\rm orb}$ & 3 \\
    \ldelim\{{3}{23mm}[M$_{\rm donor} \leq 0.07 \rm M_{\odot}$] & P$_{\rm orb} \approx 80$\,min & 2 \\
    & P$_{\rm orb} > 85$\,min & 1\\
    & P$_{\rm orb} > 90$\,min & 0 \\
    \ldelim\{{2}{23mm}[M$_{\rm donor} > 0.07 \rm M_{\odot}$] & P$_{\rm orb} \approx 80$\,min & 1 \\
    & P$_{\rm orb} > 85$\,min & 0 \\
    \ldelim\{{2}{11mm}[No mass] & P$_{\rm orb} \leq 90$\,min & 1 \\
    & P$_{\rm orb} > 90$\,min & 0 \\
    \noalign{\smallskip}
    \hline
    \noalign{\smallskip}
    \multicolumn{2}{c|}{\textbf{WD TEMPERATURE}} & \\
    \multicolumn{2}{c|}{T$_{\rm eff} \leq 12500$\,K$^4$} & 3 \\
    \multicolumn{2}{c|}{14000\,K $<$ T$_{\rm eff} \leq 14000$\,K} & 1 \\
    \multicolumn{2}{c|}{T$_{\rm eff} > 14000$\,K} & 0 \\
    \noalign{\smallskip}
    \hline
    \noalign{\smallskip}
    \multicolumn{2}{c|}{\textbf{GAIA VARIABILITY}} & \\
    \multicolumn{2}{c|}{G$_{\rm var} \leq 0.2$} & 3 \\
    \multicolumn{2}{c|}{0.2 $<$ G$_{\rm var} \leq 0.3$} & 1 \\
    \multicolumn{2}{c|}{G$_{\rm var} > 0.3$} & 0 \\
    \noalign{\smallskip}
    \hline
    \noalign{\smallskip}
    \multicolumn{2}{c|}{\textbf{GAIA COLORS}}   & \\
    \multicolumn{2}{c|}{In WD locus}   & 3 \\
    \multicolumn{2}{c|}{Broad WD locus}   & 2 \\
    \multicolumn{2}{c|}{Halfway to main-sequence}   & 1 \\
    \multicolumn{2}{c|}{In main-sequence}   & 0 \\
    \noalign{\smallskip}
    \hline
    \noalign{\smallskip}
    \multicolumn{2}{c|}{\textbf{SDSS COLORS}}  & \\
    \ldelim\{{2}{15mm}[$u-g \leq 0.5$] & $u-g \geq 1.25(g-r)-0.1$  & 3 \\
    & $u-g \geq 1.25(g-r)-0.4$  & 1 \\
    \multicolumn{2}{c|}{Other} & 0 \\
    \noalign{\smallskip}
    \hline
    \noalign{\smallskip}
    \multicolumn{2}{c|}{\textbf{GALEX COLORS}}  &\\
    \multicolumn{2}{c|}{FUV-NUV $\geq 1$} & 3 \\
    \multicolumn{2}{c|}{1 $>$ FUV-NUV $\geq 0.25$} & 1 \\
    \multicolumn{2}{c|}{FUV-NUV $< 0.25$} & 0 \\
    \noalign{\smallskip}
    \hline
    \noalign{\smallskip}
    \multicolumn{2}{c|}{\textbf{IR COLORS}} & \\
    \multicolumn{2}{c|}{Brown dwarf or L donor}  & 3\\
    \multicolumn{2}{c|}{M donor}  & 1\\
    %BD donor & $H-K \geq 0.4$ and $J-H \leq 0.5$ & 3\\
    %L donor & $H-K \geq 0.6$ and $J-H > 0.5$ & 3 \\
    %M donor & $0.4 \leq H-K < 0.6$ and $J-H > 0.5$ & 1 \\
    \multicolumn{2}{c|}{Other} & 0 \\
    \noalign{\smallskip}
    \hline
    \noalign{\smallskip}
    \multicolumn{2}{c|}{\textbf{IR EXCESS}} & \\
    \multicolumn{2}{c|}{No excess / After $K$-band}  & 3\\
    \multicolumn{2}{c|}{After $J$-band}  & 1\\
    %BD donor & $H-K \geq 0.4$ and $J-H \leq 0.5$ & 3\\
    %L donor & $H-K \geq 0.6$ and $J-H > 0.5$ & 3 \\
    %M donor & $0.4 \leq H-K < 0.6$ and $J-H > 0.5$ & 1 \\
    \multicolumn{2}{c|}{Other} & 0 \\
    \noalign{\smallskip}
    \hline

    \noalign{\smallskip}
 %   \multicolumn{3}{l}{\footnotesize {\it References.} $^1$ \cite{knigge2011}, $^2$ \cite{gansicke2009},}\\
  %  \multicolumn{3}{l}{\footnotesize $^3$ \cite{knigge2006}, $^4$ \cite{pala2022}}\\
    
\end{tabular}
\end{table}

\subsection{Defining the scoring system}\label{subsect:sys_scorecard}

In order to analyze the likelihood that a system is a period-bouncer, we assigned numerical scores to each individual parameter that we then combined into a final numerical score for each period-bounce candidate. The values obtained from all individual parameters considered in the scorecard (described in Sect.~\ref{subsect:def_scorecard}) and the final score are reported in Table~\ref{scorecard}. The maximum number of achievable score points is $36$, corresponding to an object that has the highest score in all $10$ parameters, and hence a final score of $100$\,$\%$. 
However, a considerable number of candidates did not have enough data to be scored in all of the parameters making direct comparison between systems difficult. To solve this problem, for each individual object we re-define the final score of $100$\,$\%$ as the maximum number of points that it would have received if it had the highest likelihood of being a period-bouncer in every parameter for which it has available data. We then calculated the percentage score as the ratio between the actual points the system has and its maximum achievable points. The objects without a percentage value in Table~\ref{scorecard} only had available information for three or less parameters, which we judged as not enough to characterize if this object is a period-bouncer. 

\subsection{Defining the scorecard parameters $\&$ their scores}\label{subsect:def_scorecard}

We present in Table~\ref{scorecardParameters} a summary of the scoring system for each parameter, and we illustrate this point system with Figs.~\ref{PorbMdonorSelectionCut} $-$ \ref{2MASSSelectionCut} in which the final score achieved by the systems is presented as a color scale, and candidates without enough information to have received a final score are %always 
presented in yellow. To further enhance the clarity, areas of likelihood are marked by different shape styles: filled circles for high likelihood (3 points), crosses for medium high likelihood (2 points), triangles for medium likelihood (1 point), and squares for low likelihood (0 points). The assignment of the scores to each parameter is described here in descending order of relevance: 

\begin{enumerate}
    \item {\it Spectral type of donor:} This parameter holds the most weight as spectroscopic detection of a late-type donor in the system is the ultimate confirmation needed to confidently classify a system as a period-bouncer. Period-bouncers are characterized as being CVs with degenerate donors, composed by a WD and either a brown dwarf or a L dwarf companion. We also take into account systems with a late M type (M5 or later) as this is indicative of a CV slightly before or just at the bouncing point \citep{knigge2011}. Even though a spectroscopic detection is the preferred confirmation method for a late-type companion in the system, we also consider photometric data that suggests the presence of a late-type donor in the system. 
    \\
    \item {\it Donor mass:} A very low donor mass is indicative of a highly evolved CV even if there is no direct detection of the donor. In both their standard and revised evolution tracks for CVs, \cite{knigge2011} obtained a post-bounce area for CVs with a donor mass lower than 0.058M$_{\odot}$, this limit is shown as the blue horizontal line in Fig.~\ref{PorbMdonorSelectionCut}. Systems with slightly higher donor masses (up to a donor mass of 0.7M$_{\odot}$ marked by the red line in Fig.~\ref{PorbMdonorSelectionCut}) would be placed right in the bounce area, meaning that these systems are just within range of being called a potential period-bouncer. When assigning a score to this parameter we considered both possible values for the donor mass if available (see Sect.~\ref{subsect:literatureValues} for more information), giving each of the two donor mass values an individual score. For the overall score assigned to a system with two reported values for the donor mass we considered the average of the two individual scores.\\
    
    %We rated the likelihood of being a period-bouncer as follows: high if both results for the donor mass based on the estimates introduced in Sect.\ref{sect:values} fulfilled $\rm M_{donor} \leq 0.058\rm M_{\odot}$ \citep{knigge2011}, covering the area below the blue line in Fig.~\ref{PorbMdonorSelectionCut}; medium high if both results fulfilled $\rm M_{donor} \leq 0.07\rm M_{\odot}$ \citep{knigge2011} with one of them being $\rm M_{donor} \leq 0.058\rm M_{\odot}$; medium low if both results fulfilled $\rm M_{donor} \leq 0.07\rm M_{\odot}$, covering the area between the blue and red line in Fig.~\ref{PorbMdonorSelectionCut}; low if both results fulfilled $\rm M_{donor} > 0.07\rm M_{\odot}$, covering the area above the red line in Fig.~\ref{PorbMdonorSelectionCut}.

    \item {\it Orbital period:} Near the period minimum the orbital period by itself is not a good diagnostic for evolution in CVs. To use $P_{\rm orb}$ as a parameter in the scorecard we consider also the score that each system was given for the donor mass. The clearest example of this are systems with very low donor masses ($\leq$ 0.058M$_{\odot}$; \citealt{knigge2011}) for which the orbital period does not have significant incidence as they are all clearly located in the post-bounce area (see filled circles in Fig.~\ref{PorbMdonorSelectionCut}). For all other systems, their scores are assigned judging their proximity to the observed period minimum of CVs ($\approx 80$\,min; \citealt{gansicke2009}). \\
    
    %We rated the likelihood of being a period-bouncer as follows: 
    %\begin{itemize}
    %    \item High likelihood from donor mass parameter: High for all orbital periods.
    %    \item Medium high likelihood from donor mass parameter: High for P$_{orb}\, \approx$ 80min \citep{gansicke2009}, covering the area to the left of the blue line in Fig.~\ref{PorbMdonorSelectionCut}; medium high for P$_{orb} >$ 85min, covering the area between the blue and red line in Fig.~\ref{PorbMdonorSelectionCut}; medium low for P$_{orb} >$ 90min \citep{knigge2006}, covering the area to the right of the red line in Fig.~\ref{PorbMdonorSelectionCut}.
    %    \item Medium low likelihood from donor mass parameter: medium high for P$_{orb}\, \approx$ 80min; medium low for P$_{orb} >$ 85min, covering the area between the blue and red line in Fig.~\ref{PorbMdonorSelectionCut}; low for P$_{orb} >$ 90min, covering the area to the right of the red line in Fig.~\ref{PorbMdonorSelectionCut}.
    %    \item Low likelihood from donor mass parameter: medium low for P$_{orb}\, \approx$ 80min, covering the area to the left of the blue line in Fig.~\ref{PorbMdonorSelectionCut}; low for P$_{orb} >$ 85min, covering the area to the right of the blue line in Fig.~\ref{PorbMdonorSelectionCut}.
    %    \item Unknown mass: medium low for P$_{orb} <$ 90min.\\
    %\end{itemize} 
    
%----------------------------------------------------------------- 
\begin{figure}
    \centering
    %trim={<left> <lower> <right> <upper>}
    \includegraphics[width=\columnwidth]{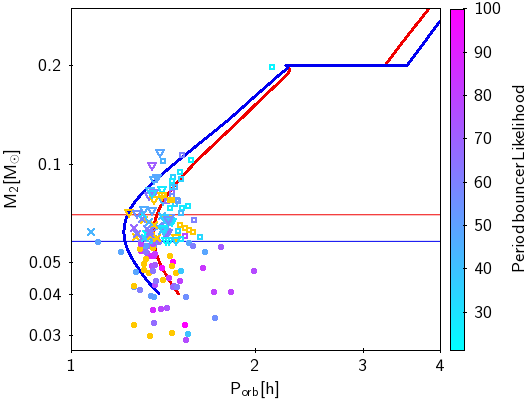}
    \caption{Donor mass as a function of orbital period for \numPorbMdonor systems in our literature catalog with available values for both orbital period and donor mass. The two \cite{knigge2011} CV evolution tracks from Fig.~\ref{PorbMdonor} are shown as reference. See text for the justification of the selection cut lines.}
    \label{PorbMdonorSelectionCut}
\end{figure}
%-----------------------------------------------------------------

    \item {\it WD temperature:} Considering that period-bouncers are the most evolved systems among CVs, a cool WD temperature (when not in outburst) is expected with temperatures $\leq$12500K (blue line in Fig.~\ref{WDteffSelectionCut}; \citealt{pala2022}). A temperature higher than 14000K (red line in Fig.~\ref{WDteffSelectionCut}) is no longer reflective of the cool WD period-bouncers are expected to exhibit, and in the rare cases when the measurement was carried out during an outbursts, it should not be considered representative of the true WD temperature.
    %We rated the likelihood of being a period-bouncer as follows: high for T$_{\rm WD} <$ 12500K \citep{pala2022}, covering the area below the blue line in Fig.~\ref{WDteffSelectionCut}; medium for T$_{\rm WD} <$ 14000K, covering the area between the blue and red line in Fig.~\ref{WDteffSelectionCut}; low for T$_{\rm WD} >$ 14000K, covering the area above the red line in Fig.~\ref{WDteffSelectionCut}.
    \\
%----------------------------------------------------------------- 
\begin{figure}
    \centering
    %trim={<left> <lower> <right> <upper>}
    \includegraphics[width=\columnwidth]{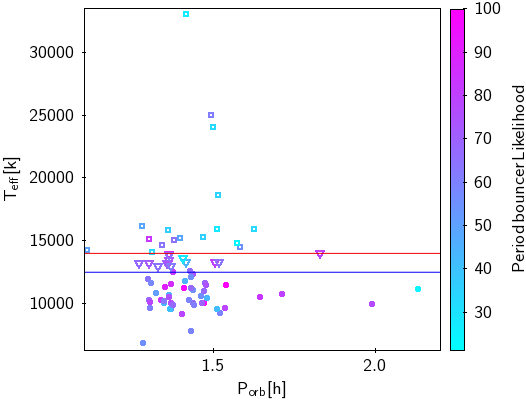}
    \caption{WD effective temperature as a function of orbital period for \numTwd systems in our literature catalog with available information. See text for the justification of the selection cut lines.}
    \label{WDteffSelectionCut}
\end{figure}
%-----------------------------------------------------------------

    \item {\it Gaia variability:} As period-bouncers are in the last stage of CV evolution they are expected to be inactive systems with very large recurrence times for both outbursts and superoutbursts in the range of $\sim$10000 days \citep{patterson2011}. For this reason, unless the period-bouncer is caught in a very rare burst episode, we expect a lack of optical variability. We used the {\it Gaia} $G$-band variability defined in Eq.~(1) of \cite{inight2023b}, and pick the limit values based on the values they obtained for WZ Sge-type CVs (see their Fig.~36). \cite{inight2023b} obtained a {\it Gaia} variability $\leq$ 0.2 for 90$\%$ of the objects categorized as WZ Sge in their sample (marked by the blue line in Fig.~\ref{GaiaVarSelectionCut}), and a {\it Gaia} variability $\leq$ 0.3 for 98$\%$ of the objects categorized as WZ Sge in their sample (marked by the red line in Fig.~\ref{GaiaVarSelectionCut}). Similar to the WD temperature, this parameter can be affected if the system was undergoing an outburst at the moment of the measurement. For this reason, we also calculated the $G$-band variability for {\it Gaia}-DR2 and assigned a score only for the candidates that received the same score points in {\it Gaia}-DR2 and {\it Gaia}-DR3, ensuring that their low or high variability is consistent through time and not reflective of sporadic events. 
    %\textcolor{orange}{\bf BS-comment: How can there be persistent high variability?} \textcolor{red}{\bf there are sub-types of dwarf novae (like ER UMa and Z Cam and even some SU UMa) that are characterized by high mass accretion rates that cause short superoutburst cycles meaning that they go from quiescence to outburst back to quiescence in timescales of 20 to 50 days} MISSING
    \\
%----------------------------------------------------------------- 
\begin{figure}
    \centering
    %trim={<left> <lower> <right> <upper>}
    \includegraphics[width=\columnwidth]{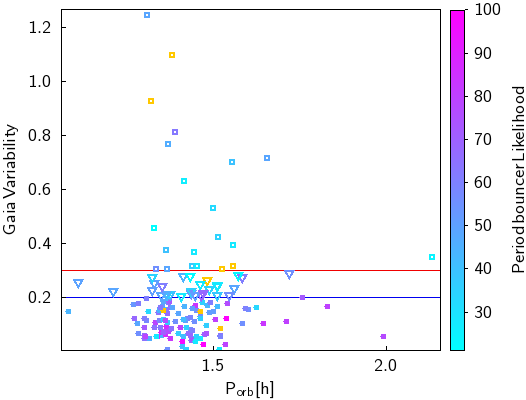}
    \caption{Gaia variability as a function of orbital period for \numGaia systems in our literature catalog with available information. See text for the justification of the selection cut lines.}
    \label{GaiaVarSelectionCut}
\end{figure}
%-----------------------------------------------------------------

    \item {\it Gaia colors:} In optical wavelengths the highly evolved secondary of period-bouncers is not expected to be observable. Therefore, these systems will appear almost identical to an isolated WD \citep{santisteban2018}. We divided Fig.~\ref{GaiaSelectionCut} in the following sectors: a WD locus limited by the blue line ($M_G \geq 2.95 \times (G_{BP}-G_{RP}) + 10.83$; \citealt{jimenez2018}), a more broadly defined WD locus limited by the blue and red lines ($M_G \geq 5 \times (G_{BP}-G_{RP}) + 6$ ; \citealt{gentile2021}, an area close but still below the main-sequence limited by the red and green lines ($M_G \geq 4.2 \times (G_{BP}-G_{RP}) + 2.5$), and a main-sequence limited by the green line.
    \\
%----------------------------------------------------------------- 
\begin{figure}
    \centering
    %trim={<left> <lower> <right> <upper>}
    \includegraphics[width=\columnwidth]{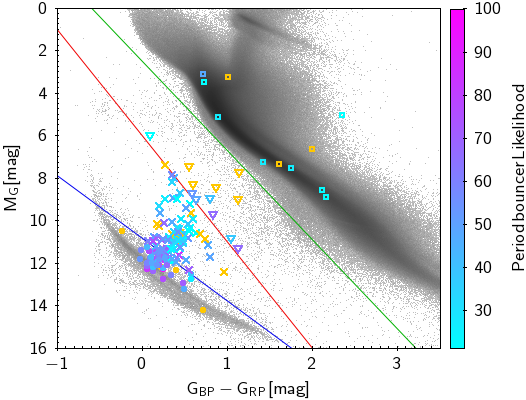}
    \caption{{\it Gaia} color-magnitude diagram showing the position of the \numGaiaHR systems in our literature catalog with available information. The {\it Gaia}-DR3 sources with \cite{bailer2021} distances and a parallax error less than 1$\%$ of the parallax value are shown in gray as a reference. See text for the justification of the selection cut lines.}
    \label{GaiaSelectionCut}
\end{figure}
%-----------------------------------------------------------------

    \item {\it SDSS colors:} We aimed to select areas where isolated WDs are expected to be found as the preferential location for period-bouncers. \cite{inight2023b} plotted a color-color diagram for CVs with SDSS data and reliable photometry (see their Fig.~16) which shows the location of different CV types. We selected the areas where the majority of the \cite{inight2023b} WZ Sge-type systems were found with 80$\%$ of them located between the blue horizontal line ($\rm u-g\leq0.5$) and the blue diagonal line ($\rm u-g\geq1.25(g-r)-0.1$) represented by the filled circles in Fig.~\ref{SdssSelectionCut}. 90$\%$ of WZ Sge-type CVs are located between the blue horizontal line and the red diagonal line ($\rm u-g\geq1.25(g-r)-0.4$) represented by the triangles in Fig.~\ref{SdssSelectionCut}.
    \\
%----------------------------------------------------------------- 
\begin{figure}
    \centering
    %trim={<left> <lower> <right> <upper>}
    \includegraphics[width=\columnwidth]{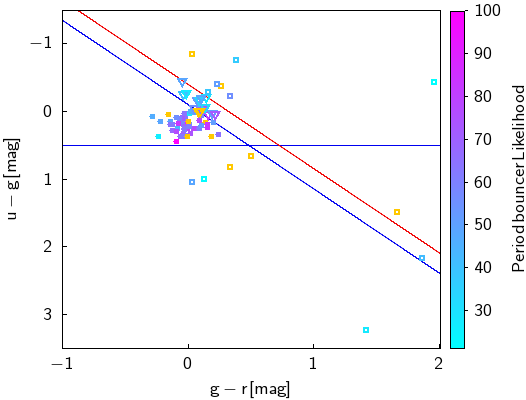}
    \caption{SDSS color-color diagram showing the position of the \numSDSS systems in our literature catalog with available information. See text for the justification of the selection cut lines.}
    \label{SdssSelectionCut}
\end{figure}
%-----------------------------------------------------------------

    \item {\it UV colors:} For systems in quiescence, GALEX colors are a sensitive probe of the effective WD temperature. \cite{patterson2011} plotted a sample of dwarf novae (see their Fig.~2) and determined areas where period-bounce candidates are most likely to be found. The area occupied exclusively by period bounce candidates is limited by $\rm FUV-NUV\geq1$ (blue line in Fig.~\ref{GalexSelectionCut}), while the area up to $\rm FUV-NUV\geq0.25$ (red line in Fig.~\ref{GalexSelectionCut}) is populated by period-bounce candidates and other types of CVs. Above the red line in Fig.~\ref{GalexSelectionCut}, \cite{patterson2011} found no period-bounce candidates.
    \\
%----------------------------------------------------------------- 
\begin{figure}
    \centering
    %trim={<left> <lower> <right> <upper>}
    \includegraphics[width=\columnwidth]{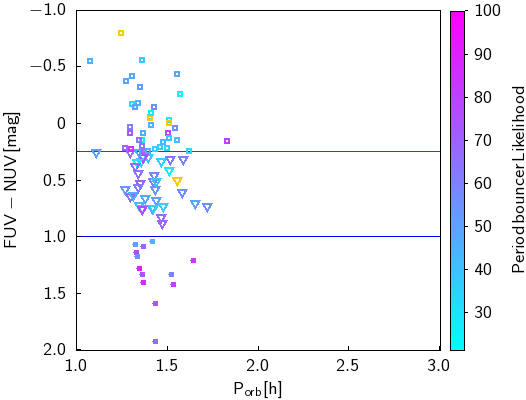}
    \caption{FUV-NUV color as a function of orbital period for \numGalex systems in our literature catalog with available information. See text for the justification of the selection cut lines.}
    \label{GalexSelectionCut}
\end{figure}
%-----------------------------------------------------------------
    \item {\it IR colors:} Several regions in the near-IR color-color diagram were identified by \cite{littlefair2003} depending on the donor type that the population of CVs had (see their Fig.~2). The right part of Fig.~\ref{2MASSSelectionCut} is populated by systems with donors identified as either brown dwarfs (bottom right) or L-type (top right), while M-type donors are found towards the center (see triangles in Fig.~\ref{2MASSSelectionCut}). Earlier type donors and other unidentified donor types are found towards the left part of Fig.~\ref{2MASSSelectionCut}. \cite{littlefair2003} selected systems with $\rm J-H \leq 3(H-K)-0.8$ (blue line in Fig.~\ref{2MASSSelectionCut}) as candidates for having a degenerate donor.
    \\
%----------------------------------------------------------------- 
\begin{figure}
    \centering
    %trim={<left> <lower> <right> <upper>}
    \includegraphics[width=\columnwidth]{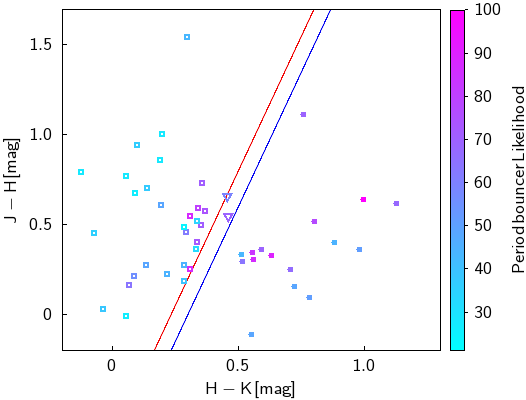}
    \caption{IR color-color diagram showing the position of the \numIR systems in our literature catalog with available information. See text for the justification of the selection cut lines.}
    \label{2MASSSelectionCut}
\end{figure}
%-----------------------------------------------------------------

    \item {\it IR excess:} Throughout the evolution of a CV, the companion star looses mass such that an originally early-type donor, with larger IR contributions, develops into a late-type or degenerate donor that barely contributes to the IR emission of the system. Due to the evolved nature of the secondary in period-bouncers, this very low mass donor is expected to appear as a very slight or no excess in the IR. We constructed SEDs for the \numSED candidates that have enough photometric data, and compared them to WD models of spectral type DA with pure hydrogen atmosphere \citep{koester2010} using the Virtual Observatory SED Analyzer (VOSA, \citealt{bayo2008}). The main takeaway from this comparison is the wavelength at which the IR excess sets in. From the population of \numPB confirmed period-bouncers we can establish that this type of CVs, especially the systems that have evolved back to periods larger than 90\,min, presents IR excess (from the donor) that starts at relatively long wavelengths associated with the $K$-band or the WISE bands, if they present an excess at all. Systems close or at the bounce point will have an excess starting at shorter wavelengths associated with the $J$-band \citep{owens2023}. Pre-bounce CVs are characterized by larger excess starting at shorter wavelengths \citep{girven2011}. In Appendix~\ref{appen:SED} we show three SEDs that are representative of the different types of IR excess described above. 
    %\textcolor{orange}{\bf BS-comment: We shall look together at some SEDs. What means "other" in the scorecard table? Is it more or less excess? Should "no excess" sources be the best PB candidates? Where in Owens+23 did you find that systems close to the PB have excess at short lambda? (I had only a very quick look at the paper)}
   %\textcolor{red}{We can also make use of the binary fit tool from VOSA that gives the temperature of the secondary (which can be associated with a spectral type). Tests that I have done with these tool give similar donor temperatures to the few ones that are available in literature.}
\end{enumerate}

In Table~\ref{scorecard}, we show the scorecard ratings for the period-bounce candidates in the literature catalog. \numNoInfo of the \numTotal systems had information for only three or less parameters. As explained above, we did not assign a score to them and we exclude them from the following analysis.

\begin{table}
    \centering
    \caption{Scorecard for the period-bounce candidates in the literature catalog. See text for explanations on the different parameters and scores. The full table is available in the online version.}
    \label{scorecard}
    \begin{tabular}{c|cccccccccc|c} 
    \hline
    \noalign{\smallskip}
    System  & \multicolumn{10}{c|}{Parameters from Sect.~\ref{sect:scorecard}} & $\%$  \\
    %\noalign{\smallskip}
    & 1 & 2 & 3 & 4 & 5 & 6 & 7 & 8 & 9 & 10 &\\
    \noalign{\smallskip}
    \hline
    \noalign{\smallskip}
    V379\,Vir  & 9 & 3 & 3 & 3 & 3 & 3 & 3 & 3 & 3 & 3 & 100 \\
    SDSS 1514  & 7 & 0 & 0 & 3 & 1 & 2 & 3 & 3 & 3 & 3 & 69 \\
    SDSS 1250  & 3 & 0 & 0 & 3 & 3 & 3 & 3 & 3 & 3 & 3 & 67 \\
    \vdots & \vdots & \vdots & \vdots & \vdots & \vdots & \vdots & \vdots &\vdots & \vdots & \vdots &\vdots \\ 
    EF Eri     & 0 & 0 & 1 & 3 & 3 & 3 & - & 1 & 0 & 1 & 37 \\
    V1258 Cen  & 0 & 0 & 0 & 3 & 0 & 1 & - & - & 0 & 1 & 17 \\
    \hline
\end{tabular}
\end{table}

Using the final scores of the \numPB systems that have been confidently categorized as period-bouncers and have a spectroscopically confirmed late-type donor (see Table~\ref{confirmedPB}), we establish that an object has a high likelihood of being a period-bouncer if its final score is higher than 60$\%$, considering that the confirmed period-bouncer with the lowest score is V406 Vir with 64$\%$. We further divided the remaining candidates as having: medium-high likelihood (between 45$\%$ and 60$\%$), medium-low likelihood (between 30$\%$ and 45$\%$) and low likelihood (lower than 30$\%$). With scores of 17$\%$ for V1258 Cen and 37$\%$ for EF Eri these systems are correctly categorized as objects with low and medium-low likelihood of being a period-bouncer respectively. This confirms that our scorecard is functioning satisfactorily as it not only returns high final scores for already confirmed period-bouncers, but it also returns low values for potential period-bouncers that have been previously discarded as good candidates. Additionally, this result shows that a system can not be classified as a period-bouncer with only one parameter, and that a multi-wavelength approach is necessary to produce a confident classification.
%\textcolor{orange}{\bf BS-comment: The reader has to be reminded what is the point about V1258 Cen. For how many parameters does it have a score? (I'm not entirely convinced that we should keep this system.)}

The differentiation of the candidates according to their final score results in \numHaMH candidates with a high or medium high likelihood of being a period-bouncer, this subsample forms a group of strong period-bounce candidates. Considering that \numPB of these candidates have already been confidently categorized as period-bouncers in the literature, this means that there are 80 strong period-bounce candidates present in the literature without having been classified as such. Final confirmation of these systems as period-bouncers would greatly contribute to the sample size for this still under-represented class of CVs. 

\section{eROSITA data}\label{sect:erosita}

To this day, eROSITA has carried out four full-sky surveys, named eRASS\,1 to eRASS\,4. Source catalogs from eRASS data are produced at Max Planck Institut für extraterrestrische Physik (MPE) in Garching, Germany, with the eROSITA Science Analysis Software System (eSASS) described by \cite{brunner2022}. These catalogs comprise all eRASS sources in the western half of the sky in terms of Galactic coordinates (Galactic longitude $l \geq 180^\circ$), which is the sky area with German data rights. 
Out of the \numTotal sources in the literature catalog 80 are located in the German eROSITA sky, 8 of which are confirmed period-bouncers listed in Table~\ref{confirmedPB}. 

\subsection{X-ray parameters for the literature sample}\label{subsect:erositaTable}

To obtain the highest sensitivity for detecting the presumably faint sources from our literature catalog, we used the merged catalog eRASS:3 which was generated from summing data from the first three all-sky surveys. The latest version of the eRASS:3 catalog available to us in December 2023 was produced with the data processing version 020\footnote{The source catalog used in our work is\\   all{\textunderscore}s3{\textunderscore}SourceCat1B{\textunderscore}221007{\textunderscore}poscorr{\textunderscore}mpe{\textunderscore}clean.fits (for eRASS:3).}. Source detection was performed in this catalog for a single eROSITA energy band, $0.2-2.3$\,keV.
%\textcolor{orange}{\bf BS-comment: You have to justify why you don't use eRASS:4; e.g. because when the work was started it was not yet available.} 

We choose to work with this eRASS:3 catalog as this was the version used to compile a catalog of CV candidates, which we use as a point of comparison with our work. This catalog of CV candidates (Schwope et al., in prep.) was produced from matching eRASS:3 and {\it Gaia} sources using \texttt{NWAY}, a software for probabilistic cross-matching of catalogs \citep{salvato2018}. It makes use of a Bayesian prior. This was trained on a set of $624$ known CVs with well-known X-ray and optical properties. The optical properties used in characterizing the sample are brightness, color, coordinates, parallax, proper motion, and variability. The X-ray properties used are position and flux, and thus implicitly the optical to X-ray flux ratio. The \texttt{NWAY} match applied to eRASS:3 using the CV prior revealed 11113 candidates with a CV probability $>50$\%, which are used in this paper for comparison to the period-bouncer sample. Full details on the construction of the eRASS:3 sample of CV candidates will be given by Schwope et al. (in prep.).

We corrected the coordinates of the period-bounce candidates in the literature catalog to the mean observing date of eRASS:3 using their {\it Gaia}-DR3 proper motions. We then matched them with the eRASS:3 catalog, allowing for a maximum separation of $30^{\prime\prime}$ and enforcing for the separation between optical and X-ray coordinates the condition sep$_{\rm ox} < 3 \times RADEC\_ERR$, where $RADEC\_ERR$ is the positional error of the X-ray coordinates in units of arcseconds. This way we found that \numSThree of the \numGS period-bounce candidates from the literature catalog that are in the German eROSITA sky are detected in the eRASS:3 catalog. 

This process was performed allowing for all possible matches within $30^{\prime\prime}$, however each of the \numSThree detected period-bounce candidates had only one eRASS:3 match within that radius. We then carried out a visual inspection using ESAsky\footnote{\url{https://sky.esa.int/}} in a $30^{\prime\prime}$ radius region around the X-ray source to assure that there where no other potential optical counterparts. Out of the \numSThree period-bounce candidates detected in the eRASS:3 catalog, \numSThreeApproved have no other optical source closer to the eRASS X-ray position than our target and can, therefore, be confidently categorized as a correct match. Two additional eRASS:3 detections were confirmed as the correct match thanks to previous X-ray detections ({\it XMM-Newton} and {\it Chandra}) clearly associated with our targets. Out of the remaining 4 eRASS:3 detections, for one of them the visual inspection revealed an eRASS source at a sep$_{\rm ox}$ that is larger than the maximum of $3 \times RADEC\_ERR$ that we allowed in the automatic match. %\textcolor{orange}{\bf BS-comment: Is that what you want to say? At how many arcsec? Could it still be the right source? Is there an XMM/Chandra detection?}\textcolor{red}{I meant the sep < 3xRADEC condition we enforced. The separation is 15arcsec ($\sim$2arcsec larger than the RADEC condition allows) and there are no closer Gaia sources. There are no other X-ray detections around this target so we can not verify it this way.} 
For the last three there is another object closer to the eRASS source than our target, such that the X-ray source could not be securely associated to the period-bounce candidate. We present in Table~\ref{eROSITAcat} the \numSThree sources with an eROSITA detection, where we do not report X-ray parameters for the 4 sources without a reliable association. These 4 sources are not considered in the following analysis.

\begin{table*}[h]
    \centering
    \begin{threeparttable}
    \caption{X-ray parameters from the eROSITA merged catalog eRASS:3 for the detected sources from the literature catalog, known as the eROSITA subsample. Values are given for the eROSITA single band (0.2-2.3\,keV).}         
    \label{eROSITAcat}                            
    \begin{tabular}{l c c c c c c c}        
    \hline              
    \noalign{\smallskip}
    Name & Detection & sep$_{\rm ox}$ & Counts & Count rate  &  Flux ({\sc apec}) &  X-ray luminosity$^1$ & Mass accretion rate\\ 
    & likelihood & [$^{\prime\prime}$] &[cts] &[cts/s] & $\times 10^{-13}$ & log($\rm L_{bol}$) & $\times 10^{-14}$ \\
     & & && & [${\rm erg\, cm^{-2} s^{-1}}$] & [erg$ s^{-1}$] & [$M_\odot \rm yr^{-1}$ ]\\
    \hline                      
    \noalign{\smallskip}
    V379 Vir$^\dagger$& 44.5& 2.2& 30.1$\pm$6.3& 0.08$\pm$0.02& 0.75$\pm$0.16& 29.5& 3.2\\
    SDSS 1250$^\dagger$& 19.9& 2.9& 16.5$\pm$4.9& 0.05$\pm$0.02& 0.46$\pm$0.14& 29.2& 1.0\\
    EF Eri& 100.3& 6.2& 55.0$\pm$8.2& 0.1$\pm$0.01& 0.86$\pm$0.13& 29.6& 2.2\\
    SDSS 1057$^\dagger$& 15.5& 6.0& 10.7$\pm$3.7& 0.05$\pm$0.02& 0.42$\pm$0.14& 30.0& 6.0\\
    SDSS 1035$^\dagger$& 5.0& 13.3& 3.8$\pm$2.1& 0.02$\pm$0.01& 0.16$\pm$0.09& 29.1& 0.7\\
    QZ Lib$^\dagger$& 107.1& 1.6& 64.6$\pm$9.2& 0.14$\pm$0.02& 1.26$\pm$0.18& 30.0& 6.1\\
    RX 1050$^*$& 131.9& 2.1& 52.3$\pm$7.7& 0.18$\pm$0.03& 1.64$\pm$0.24& 29.6& 2.6\\
    AL Com& 20.6& 7.3& 14.7$\pm$4.5& 0.05$\pm$0.01& 0.42$\pm$0.13& 30.5& 15.9\\
    EG Cnc$^*$& 14.0& 8.8& 9.4$\pm$3.5& 0.05$\pm$0.02& 0.42$\pm$0.16& 29.5& 1.1\\
    IL Leo& 595.3& 2.1& 162.4$\pm$13.4& 0.75$\pm$0.06& 6.71$\pm$0.55& 31.6& 227.9\\
    GW Lib& 182.4& 3.2& 85.8$\pm$10.1& 0.21$\pm$0.02& 1.89$\pm$0.22& 29.7& 2.8\\
    V406 Vir$^\dagger$& 83.0& 0.9& 36.5$\pm$6.5& 0.11$\pm$0.02& 1.01$\pm$0.18& 29.8& 2.4\\
    SDSS 1216$^*$& 22.4& 3.0& 15.2$\pm$4.6& 0.05$\pm$0.02& 0.46$\pm$0.14& 30.1& 8.1\\
    BW Scl$^\dagger$& 766.5& 0.7& 204.1$\pm$14.9& 0.82$\pm$0.06& 7.33$\pm$0.54& 30.1& 5.9\\
    RX 0232& 363.3& 1.7& 120.6$\pm$11.6& 0.25$\pm$0.02& 2.26$\pm$0.22& 30.3& 5.4\\
    SDSS 0903& 53.3& 4.6& 25.2$\pm$5.4& 0.11$\pm$0.02& 0.96$\pm$0.2& 30.3& 11.0\\
    SSS 1222$^\dagger$& 199.8& 1.9& 84.6$\pm$9.8& 0.22$\pm$0.03& 1.93$\pm$0.22& 30.4& 11.6\\
    SDSS 0904& 6.5& 11.5& 6.1$\pm$2.9& 0.03$\pm$0.01& 0.25$\pm$0.12& 29.6& 2.7\\
    SDSS 1219$^*$& 32.1& 2.6& 18.6$\pm$4.8& 0.06$\pm$0.02& 0.56$\pm$0.15& 29.9& 5.2\\
    OT 0600& 19.1& 4.7& 12.1$\pm$3.9& 0.04$\pm$0.01& 0.4$\pm$0.13& 30.8& 40.5\\
    HV Vir$^*$& 11.7& 11.6& 11.1$\pm$4.0& 0.03$\pm$0.01& 0.29$\pm$0.1& 29.7& 3.1\\
    KN Cet$^*$& 129.4& 1.5& 58.6$\pm$8.3& 0.12$\pm$0.02& 1.09$\pm$0.15& 30.1& 9.0\\
    OY Car& 5833.9& 0.5& 1418.7$\pm$39.4& 1.21$\pm$0.03& 10.76$\pm$0.3& 30.2& 8.9\\
    XZ Eri& 25.9& 8.1& 18.4$\pm$4.9& 0.04$\pm$0.01& 0.37$\pm$0.1& 30.0& 6.9\\
    AQ Cmi& 265.5& 1.9& 87.2$\pm$9.9& 0.35$\pm$0.04& 3.09$\pm$0.35& 31.3& 122.5\\
    AQ Eri& 1488.0& 1.3& 381.8$\pm$20.5& 0.99$\pm$0.05& 8.84$\pm$0.47& 31.4& 156.9\\
    DT Pyx& 12.8& 0.7& 9.4$\pm$3.5& 0.03$\pm$0.01& 0.3$\pm$0.11& 29.8& 4.1\\
    IK Leo$^*$& 14.0& 1.6& 11.8$\pm$4.1& 0.05$\pm$0.02& 0.47$\pm$0.16& 30.7& 33.9\\
    SDSS 0734& 70.5& 4.1& 29.2$\pm$5.8& 0.13$\pm$0.03& 1.13$\pm$0.23& 31.7& 291.5\\
    SDSS 0947& 25.3& 3.9& 13.9$\pm$4.1& 0.06$\pm$0.02& 0.57$\pm$0.17& 30.8& 40.6\\
    CRTS 1044& 38.6& 0.2& 18.0$\pm$4.7& 0.08$\pm$0.02& 0.73$\pm$0.19& 30.4& 15.2\\
    SDSS 1058& 21.4& 2.4& 11.8$\pm$3.7& 0.06$\pm$0.02& 0.52$\pm$0.16& 30.7& 31.1\\
    CRTS 1259& 7.0& 6.4& 6.4$\pm$3.0& 0.02$\pm$0.01& 0.16$\pm$0.08& 30.1& 8.3\\
    TCP 1537& 17.8& 5.8& 15.4$\pm$4.9& 0.04$\pm$0.01& 0.36$\pm$0.12& -& - \\
    KK Cnc& 11.4& 5.2& 7.4$\pm$3.1& 0.03$\pm$0.01& 0.29$\pm$0.12& 30.6& 25.6\\
    MM Hya& 18.6& 7.6& 10.8$\pm$3.7& 0.05$\pm$0.02& 0.41$\pm$0.14& 30.0& 5.8\\
    MM Sco& 117.6& 2.0& 60.1$\pm$8.6& 0.18$\pm$0.03& 1.62$\pm$0.23& 30.7& 30.1\\
    RX Vol& 269.8& 1.6& 127.2$\pm$12.2& 0.1$\pm$0.01& 0.93$\pm$0.09& 31.7& 297.1\\
    RZ Lmi& 34.3& 4.6& 18.4$\pm$4.8& 0.08$\pm$0.02& 0.71$\pm$0.19& 30.8& 36.8\\
    V436 Cen& 2651.2& 1.0& 658.3$\pm$26.7& 1.62$\pm$0.07& 14.41$\pm$0.59& 30.8& 53.6\\
    V591 Cen& 15.6& 6.3& 13.9$\pm$4.4& 0.03$\pm$0.01& 0.27$\pm$0.09& 31.1& 86.7\\
    VX For& 41.2& 3.3& 27.5$\pm$6.0& 0.04$\pm$0.01& 0.35$\pm$0.08& 30.3& 11.5\\
    RX 0154& 3488.2& 0.9& 785.3$\pm$29.1& 1.8$\pm$0.07& 16.04$\pm$0.59& 31.5& 229.7\\
    RX 0600& 445.8& 1.6& 141.6$\pm$12.6& 0.34$\pm$0.03& 3.05$\pm$0.27& 31.9& 630.1\\
    ASASSN -14jf& 69.0& 1.6& 57.2$\pm$8.9& 0.02$\pm$0.0& 0.22$\pm$0.03& 30.3& 12.6\\
    ASASSN -17el$^*$& 188.2& 2.6& 93.0$\pm$10.6& 0.13$\pm$0.01& 1.16$\pm$0.13& 30.4& 17.0\\
    CRTS 0522$^*$& 25.2& 5.2& 20.3$\pm$5.3& 0.03$\pm$0.01& 0.28$\pm$0.07& 30.5& 23.4\\
    \noalign{\smallskip}
    \hline 
    \noalign{\smallskip}
    OT 1112 & - & 7.1 & - & - & - & - & - \\
    PNV 1714 & - & 2.0 & - & - & - & - & - \\
    V1258 Cen & - & 15.1 & - & - & - & - & - \\
    FL TrA & - & 5.5 & - & - & - & - & - \\
    \noalign{\smallskip}
    \hline 
    \end{tabular}
    \begin{tablenotes}
        \small
        \item $^1$ The X-ray luminosity refers to the bolometric band from 0.1-12\,keV. $^\dagger$ System is a confirmed period-bouncer. $^*$ System was selected by us as a strong period-bounce candidate.
    \end{tablenotes}
    \end{threeparttable}
\end{table*}

We note in passing that of the \numSThreeApprovedP systems with safe eRASS:3 detection 31 are detected when using exclusively the first all-sky survey catalog eRASS\,1\footnote{The source catalog used in our work is the 020 version of\\  all{\textunderscore}e1{\textunderscore}SourceCat3B{\textunderscore}221031{\textunderscore}poscorr{\textunderscore}mpe{\textunderscore}clean.fits (for eRASS1).}. The use of the combined catalog from three surveys, thus, constitutes a  significant improvement, which is not unexpected as our targets are faint and substantially benefit from deeper X-ray exposure.

We then proceeded with our analysis for the \numSThreeApprovedP systems with safe eRASS:3 detection, referred to as "eROSITA subsample", which includes all 8 confirmed period-bouncers located in the German eROSITA sky. In Table~\ref{eROSITAcat} we present their X-ray parameters from eROSITA. In cols.~2-5 we present the separation sep$_{\rm ox}$, the detection likelihood value (the eRASS:3 catalog has a minimum detection likelihood of 5.0), the number of net source counts, and the count rate. The latter two refer to the $0.2-2.3$\,keV band used in eRASS:3. The catalogue uses a power-law model in order to obtain the flux from the count rate. This is not suitable for CVs which are characterized by a thermal plasma. Therefore, we converted the catalog flux from a power-law model to an {\sc apec} model using a conversion factor of $1.04$ (see \citealt{munoz2023} for a more detailed description). Using the {\sc apec} flux (col.~6 of Table~\ref{eROSITAcat}) together with the {\it Gaia} DR3 distances given by \cite{bailer2021} we obtained the X-ray luminosity, which was converted into bolometric X-ray luminosity ($L_{\rm x,bol}$) in the band 0.1-12\,keV (given in col.~7) by multiplication with a factor $1.6$ \citep{munoz2023}. We then calculated the mass accretion rate for each system taking into consideration if it has a reported value for the WD mass (see Table~\ref{PBtable}). For those systems with a reported WD mass we used this value together with the WD radius obtained with the \cite{nauenberg} mass-radius relation. For the systems without a WD mass we used 0.8M$_\odot$, which is the mean mass of WDs in CVs \citep{pala2022}, and the corresponding radius of $7 \times 10^8$cm from the \cite{nauenberg} mass-radius relation. Only one system, TCP 1537, does not have a reported {\it Gaia} DR3 distance which prevents us from calculating a X-ray luminosity and mass accretion rate for this period-bounce candidate.    

\subsection{The X-ray parameter space of period-bouncers}\label{subsect:erositaSelection}

Using the final scores assigned to the period-bounce candidates (see Table~\ref{scorecard}) together with the X-ray results from eROSITA (see Table~\ref{eROSITAcat}) we can establish two new parameters, X-ray-to-optical flux ratio ($\rm F_{\rm x}/F_{\rm opt}$) and bolometric X-ray luminosity, that will aid in the future identification of highly likely period-bounce candidates from eROSITA data. 
%\textcolor{orange}{\bf BS-comment: (1) Not clear to me at this point why you introduce Fx and Lx as additional parameters for PB status. Will you define a score for them. Maybe it becomes clear when you have completed this section? (2) Also: Why should Fx and Lx be two (independent) parameters? (3) For defining X-rays as a PB criterion you do not need the info from the scorecard (Table 4). }
A possible third new parameter, the mass accretion rate ($\rm \dot{M}_{\rm acc}$), could be used for cases with known WD mass,
considering that using the mean WD mass and radius gives a distribution indistinguishable from the one obtained using the X-ray luminosity. As less than half of the period-bounce candidates with eROSITA data have an individually determined WD mass, $\rm \dot{M}_{\rm acc}$ would not yield additional information on the sample and we do not use it to define selection cuts for period-bouncers. 

\begin{figure}
    \centering
    %trim={<left> <lower> <right> <upper>}
    \includegraphics[width=\columnwidth]{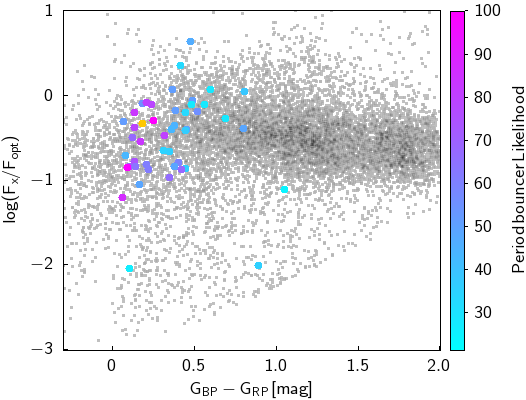}
    \caption{X-ray-to-optical flux ratio as a function of {\it Gaia} colors showing the position of the \numSThreeApprovedPdistance systems in our literature catalog found in eRASS:3. We show in grey as reference a population of CV candidates found in eRASS:3.}
    \label{XrayOpticalS3}
\end{figure}

\begin{figure}
    \centering
    %trim={<left> <lower> <right> <upper>}
    \includegraphics[width=\columnwidth]{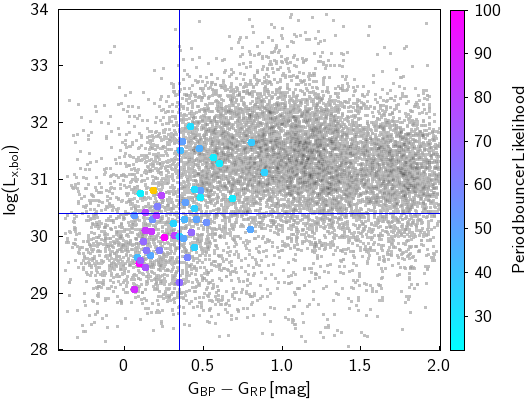}
    \caption{Bolometric X-ray luminosity as a function of {\it Gaia} colors showing the position of the \numSThreeApprovedPdistance systems in our literature catalog found in eRASS:3. We show in grey as reference a population of CV candidates found in eRASS:3. See text for the justification of the selection cut lines.}
    \label{bolLuminosityS3}
\end{figure}

In Figs.~\ref{XrayOpticalS3} and \ref{bolLuminosityS3} we have placed the eROSITA subsample in two diagrams combining X-ray and optical data. We compare the position of this sample with the overall catalog of eROSITA selected CV candidates (Schwope et al., in prep) introduced in Sect~\ref{subsect:erositaTable}.

Fig.~\ref{XrayOpticalS3} presents the X-ray-to-optical flux ratio versus {\it Gaia} color, where the majority of both the overall eROSITA CV candidate population and our eROSITA subsample display $-1 \lesssim \rm log(F_x/F_{opt}) \lesssim 0$. This is to be expected in systems with low mass transfer rate including different CV types like dwarf novae, polars and intermediate polars which make up the bulk of the observed CV population. Eventhough, this parameter will not be used on its own to determine new period-bounce candidates, it serves as an important check that the system does not display very small values of X-ray-to-optical flux ratio, that would indicate a system with a very high mass transfer rate (e.g., nova-like variables) and thus disqualify the system from being a period-bouncer. 
%\textcolor{orange}{\bf BS-comment: (1) Discuss the origin of the "narrow stripe"  $-->$ Gaia color is blue because WD dominated. More interesting is, however the X-ray range. However, it seems that in terms of Fx/Fopt the PBs behave similar to the rest of the CVs. Is that expected or surprising? (2) in the Lxbol diagram the PB candidates show (not surprisingly) fainter Lx than the bulk of the optically redder CVs. Annie's sample seems to have two groups of objects: the optically redder main part of the CVs, and a population of blue and X-ray faint systems. Could all of Annie's objects that overlap with your literature sample be PBs ?? Axel will have to comment on this. (3) Could some plot together with Annie's sample that involves distance be interesting? e.g. Lx vs distance.}  

%----------------------------------------------------------------- 
%-----------------------------------------------------------------

Fig.~\ref{bolLuminosityS3} shows the bolometric X-ray luminosity versus {\it Gaia} color in which we identify a region preferably dominated by highly likely period-bouncers. In this diagram, the highly likely period-bouncers from the eROSITA subsample are considerably differentiated from the overall eROSITA CV candidate population displaying a lower bolometric X-ray luminosity of $L_{\rm x,bol} \approx 10^{30}$ erg/s. Together with their relatively blue {\it Gaia} color $L_{\rm x,bol}$ proves to be a powerful tool for identifying new period-bounce candidates from eROSITA data, as, compared to most other CV candidates, probable period-bouncers present both bluer colors (because they are WD dominated) and lower X-ray luminosities (because of their low mass accretion rate). Applying selection cuts based on the upper limits for the X-ray luminosity (log(L$_{\rm x,bol})\leq$ 30.4 [erg/s]) and {\it Gaia} color ($G_{BP}-G_{RP}\leq$ 0.35) exhibited by systems from our catalog that have already been confirmed as being period-bouncers (see lower left rectangle in Fig.~\ref{bolLuminosityS3}) separates $971$ CV candidates from the rest of the overall eROSITA CV candidate population. Additionally, we checked that they are low mass transfer rate systems using the X-ray-to-optical flux ratio ($-1.21 \leq \rm log(F_x/F_{opt}) \leq 0$) exhibited by the confirmed period-bouncers in our catalog, finding  that $775$ of them fulfill both X-ray criteria defined for confirmed period-bouncers from eROSITA.

%----------------------------------------------------------------- 
%-----------------------------------------------------------------

\begin{figure}
    \centering
    %trim={<left> <lower> <right> <upper>}
    \includegraphics[width=\columnwidth]{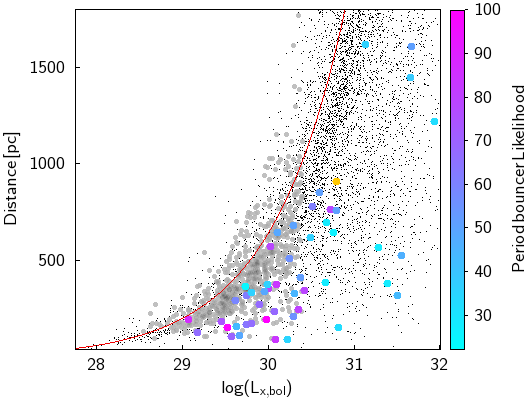}
    \caption{Distance from \cite{bailer2021} %as a function of the 
    versus 
    bolometric X-ray luminosity showing the position of the \numSThreeApprovedPdistance systems in our literature catalog detected in eRASS:3 with respect to the average eRASS:3 sensitivity limit marked by the red line. We show in black the population of CV candidates found in eRASS:3, and in grey a subsample of 775 CV candidates of them that we have selected as period-bounce candidates.}
    \label{distaceLogl}
\end{figure}

We show in Fig.~\ref{distaceLogl} the relationship between X-ray luminosity and distance. The red line gives the lower limit for the eRASS:3 luminosity calculated for the limiting X-ray flux of $2 \times 10^{-14}\,{\rm erg/cm^2/s}$ \citep{munoz2023}. Two things are immediately apparent from Fig.~\ref{distaceLogl}. First, at a given luminosity period-bounce candidates tend to have smaller distances than the majority of CV candidates.  This is expected as period-bouncers are intrinsically faint. Secondly, for a given distance, our period-bounce candidates are among the sources with the highest X-ray luminosity, probably due to (non X-ray) selection effects of the literature catalogs they were pulled from. This means that there might be a significant population of period-bouncers at short distances with X-ray luminosities lower than the period-bounce candidates identified in this work but higher than the eRASS:3 limit. These systems have yet to be singled out and studied.
%\textcolor{orange}{\bf BS-comment: The upper envelope of the bulk of the sample (of Annie) should trace the eRASS:3 flux sensitivity limit: $f_{x,lim} = L-x * 4 \pi d^2)$ where $f_{x,lim}$ is a single number for all stars, something like $2 10^{-14} erg/cm^2/s$. I think in your previous paper we have cited some value for the limiting flux of the eRASS:3 cat. Can you overplot this line, solving the above equation for the distance? For a given distance your high-prob PBs are among the ones with the highest Lx. This is opposite to what I naively would have expected. Can you make the same plot where as color code for your literature sample you use their DETLIKE value or ML\_RATE? The ones with higher values in these parameters should for a given distance be more on the right. If you have these two parameters for Annie's sample you can also give the same color code to them. One suspicion: Your literature catalog is not X-ray selected but all the other scorecard parameters are probably more easy to measure for brighter objects (at a given distance). However, that would mean that there are many PBs among annie's sample at small distances and low Lx, still do be found. Can you select from Fig.10 all of Annie's stars in the lower left of the boxes you defined, and highlight them in Fig.11? }

%----------------------------------------------------------------- 
%-----------------------------------------------------------------

In order to select this missing population of period-bouncers in eROSITA data we should look into the $775$ selected CV candidates that are systems with low mass transfer rates as well as fulfill the selection cuts shown in Fig.~\ref{bolLuminosityS3}. These systems are shown as grey dots in Fig.~\ref{distaceLogl} located fairly close to the luminosity limit for eRASS:3 (see red line in Fig.~\ref{distaceLogl}).

\section{Conclusions}\label{sect:conclusions}

One of our goals in this paper was to establish the X-ray properties of the class of period-bounce CVs, specifically using new data from the eROSITA all-sky surveys. We explored the eRASS:3 catalog with a sample of \numPB confirmed period-bouncers and 175 additional candidates that we compiled from the literature. We established two selection cuts based on the X-ray-to-optical flux ratio and the X-ray luminosity observed from the already confirmed period-bouncers. Five candidates with high likelihood of being a period-bounce system according to the multi-parameter scorecard defined by us are found within these X-ray selection cuts, meaning that they appear very similar to known period-bouncers in our multi-wavelength study including X-rays. Based on this we can confidently suggest 5 systems in our literature catalog as new period-bouncers. 4 of these 5 systems (RX 1050, EG Cnc, SDSS 1216, and HV Vir) had already been suggested as potential period-bouncers, however a confident classification was not evident to us from the literature. The remaining system, SDSS 1219, was only mentioned in the literature as a WZ Sge-type system. None of the new period-bouncers have a detected late-type donor mainly due to the lack of in-depth studies of these sources. Future detailed spectroscopic studies will shed more light on their status as period-bouncers.
%Nine candidates with high likelihood of being a period-bounce system according to the multi-parameter scorecard defined by us are found within these X-ray selection cuts, meaning that they appear very similar to known period-bouncers in our multi-wavelength study including X-rays. Based on this we can confidently suggest 9 systems in our literature catalog as new period-bouncers. 5 of these 9 systems (RX 1050, EG Cnc, SDSS 1216, HV Vir and KN Cet) had already been suggested as potential period-bouncers, however a confident classification was not evident to us from the literature. The remaining 4 systems (SDSS 1219, IK Leo, ASASSN -17el and CRTS 0522) were only mentioned in the literature as short-period CVs or WZ Sge-type systems. None of the new period-bouncers have a detected late-type donor mainly due to the lack of in-depth studies of these sources. Future detailed spectroscopic studies will shed more light on their status as period-bouncers.

This new addition of confirmed period-bouncers represents an increment in population of $\sim 30 \%$, bringing the number of this elusive class of CVs to $22$, establishing eROSITA as a powerful tool for the characterization and identification of period-bouncers. However, despite this substantial increase of the known population of period-bounce CVs, it still remains well below the numbers expected by theoretical models. We foresee that a further exploitation of eROSITA data might boost the population number to the predicted levels.
%\textcolor{red}{Comment on confirmation of highly likely candidates? 9 candidates in our catalog with high likelihood from score card also fulfill the Xray cuts from flux and luminosity. 5 of them have already been suggested as PB candidates but a definitive categorization has not been made probably because of the lack of a spectroscopic late-type donor. The other 4 have only been referred in literature as WZ Sge. Can we confidently say that this scorecard+Xray information is enough to produce such a confirmation?}
%In order to 

In an exploratory study we have tentatively identified a $500$\,pc volume-limited sample of potential period-bouncers from the overall CV population presented in a new eROSITA-selected catalog (Schwope et al., in prep.). The $500$\,pc boundary is motivated by the limitation on the accuracy of distance measurements that can be achieved with {\it Gaia} for faint objects (see Fig.~\ref{distaceBJplx}) and the 
eRASS sensitivity limit (Fig.~\ref{distaceLogl}). 
%detection limit for period-bouncers in eROSITA. 
%
%In Sect.~\ref{subsect:distance} we noticed a trend with distance that seems to affect specially faint objects like period-bouncers. The greater the distance for a faint object, the larger the fractional error of the distance (see Fig.~\ref{distaceBJplx}). For far away systems this makes unreliable not only their distance but also several additional parameters that depend on this. With this motivation we selected a 500pc volume-limited subsample, 

First, in our catalog of period-bounce candidates, 77 of the 81 systems within $500$\,pc have a fractional 
distance error smaller than 20$\%$, including all \numPB confirmed period-bouncers from Table~\ref{confirmedPB} and all 5 of the new eROSITA-confirmed period-bouncers. Secondly, when we take the mean X-ray luminosity of the 8 confirmed period bouncers detected in eRASS:3 ($\rm log(L_{x,bol}) = 29.74$ [erg/s]) as a typical X-ray luminosity for this class of objects the average eRASS:3 flux limit yields a distance limit of $480$\,pc. A rough distance limit of $500$\,pc seems, thus, appropriate for a meaningful period-bouncer population study.

In the near future we will seek to systematically uncover the $500$\,pc sample of period-bouncers, through an in-depth study of both the $81$ candidates from the catalog presented in this work and the $543$ new eROSITA CV candidates within $500$\,pc selected by us as potential period-bouncers using the X-ray selection cuts.
%Of the 81 candidates in our catalog within 500pc %only 36 are located in the German eROSITA sky, with 29 of them having X-ray data from eROSITA. %Of the remanding 7 period-bounce candidates without X-ray data we submitted 6 for an X-ray {\it XMM-Newton} proposal, as they all had final scores categorizing them as very likely period-bouncers, but have not yet been classified as such in the literature (CHECK). The last object without X-ray data is V1258 Cen which has been already categorized as not a period-bouncer. 
%The potential  confirmation of 
The confirmation or rejection of the systems from this sample will provide a benchmark for population studies of CVs around the period-bounce region.
%n additional boost to the population of period-bouncers.

%Additionally, we will also include in this volume-limited sample XX of the 1069 CV candidates that were selected using the X-ray selection cuts and that are located within 500pc.

\begin{acknowledgements}
%\textcolor{red}{Is there a special eROSTEP acknowledgment? We are citing your grant only. Maybe ask Miltos who also publishes in the DR1.} I checked several eROSTEP members and could not find an eROSTEP acknowledgment in any of their papers.
Daniela Muñoz-Giraldo acknowledges financial support from Deutsche Forschungsgemeinschaft (DFG) under grant number STE 1068/6-1. This work is based on data from eROSITA, the primary instrument aboard SRG, a joint Russian-German science mission supported by the Russian Space Agency (Roskosmos), in the interests of the Russian Academy of Sciences represented by its Space Research Institute (IKI), and the Deutsches Zentrum f\"ur Luft- und Raumfahrt (DLR). The SRG spacecraft was built by Lavochkin Association (NPOL) and its subcontractors, and is operated by NPOL with support from the Max Planck Institute for Extraterrestrial Physics (MPE). The development and construction of the eROSITA X-ray instrument was led by MPE, with contributions from the Dr. Karl Remeis Observatory Bamberg and ECAP (FAU Erlangen-N\"urnberg), the University of Hamburg Observatory, the Leibniz Institute for Astrophysics Potsdam (AIP), and the Institute for Astronomy and Astrophysics of the University of T\"ubingen, with the support of DLR and the Max Planck Society. The Argelander Institute for Astronomy of the University of Bonn and the Ludwig Maximilians Universit\"at M\"unchen also participated in the science preparation for ero. The eROSITA data shown here were processed using the eSASS/NRTA software system developed by the German eROSITA consortium. This work has made use of data from the European Space Agency (ESA) mission {\it Gaia} (\url{https://www.cosmos.esa.int/gaia}), processed by the {\it Gaia} Data Processing and Analysis Consortium (DPAC, \url{https://www.cosmos.esa.int/web/gaia/dpac/consortium}). Funding for the DPAC has been provided by national institutions, in particular the institutions participating in the {\it Gaia} Multilateral Agreement. This publication makes use of VOSA, developed under the Spanish Virtual Observatory (\url{https://svo.cab.inta-csic.es}) project funded by MCIN/AEI/10.13039/501100011033/ through grant PID2020-112949GB-I00. VOSA has been partially updated by using funding from the European Union's Horizon 2020 Research and Innovation Programme, under Grant Agreement n$^\circ$ 776403 (EXOPLANETS-A)

\end{acknowledgements}

% WARNING
%-------------------------------------------------------------------
% Please note that we have included the references to the file aa.dem in
% order to compile it, but we ask you to:
%
% - use BibTeX with the regular commands:
%   \bibliographystyle{aa} % style aa.bst
%   \bibliography{Yourfile} % your references Yourfile.bib
%
% - join the .bib files when you upload your source files
%-------------------------------------------------------------------

\bibliography{references.bib}
\bibliographystyle{aa}

\begin{appendix} %First appendix

\onecolumn
\section{Shortened Literature Catalog Table}\label{appen:litTable}

Table~\ref{PBtable_col} gives a brief description of the columns available in the complete version of our literature catalog. Selected columns showing system parameters from the literature are given in Table~\ref{PBtable}.

\begin{longtable}{c l c l}
    \caption{Content of the 66 columns in our literature catalog of period-bounce candidates, corresponding to: values obtained from the literature, photometry.}       
    \label{PBtable_col}   \\
    \hline
    \noalign{\smallskip}
    $\#$ & Name & Unit & Description \\
    \hline
    \noalign{\smallskip}
    \endhead
   % & [h] & $\rm[M_{\odot}]$ & & [pc] & [K] & $\rm [M_{\odot}]$ &  & \\
   % \noalign{\smallskip}
    1 & System & & Object name more commonly used in literature. \\
    2 & GaiaDR3 & & {\it Gaia} ID from data release 3. \\
    3 & RA & deg & Right Ascension (J200). \\
    4 & DEC & deg & Declination (J200). \\
    5 & Porb & h & Orbital period of the system. \\
    6 & Mdonor$\textunderscore$1 & M$_{\odot}$ & Donor mass. \\
    7 & Method$\textunderscore$1 & & Method used to determine the donor mass: \\
     & & & SH - Superhumps \\
     & & & EM - Eclipse modelling \\
     & & & RV - Radial velocity \\
     & & & GR - Gravitational redshift \\
     & & & SED - SED fitting \\
    8 & Mdonor$\textunderscore$2 & M$_{\odot}$ & Donor mass. \\
    9 & Method$\textunderscore$2 & & Method used to determine the donor mass. \\
    10 & SpTdonor & & Donor spectral type. \\
    11 & Method & & Method used to determine the donor spectral type: \\
    & & & S - Spectroscopic \\
    & & & P - Photometric \\
    & & & M - Assumed from mass value \\
    12 & Distance & pc & Distance to the system from \cite{bailer2021}. \\
    13 & T$\textunderscore$WD & K & WD temperature. \\
    14 & M$\textunderscore$WD & M$_{\odot}$ & WD mass. \\
    15 & WDmagnetism & & Magnetism of the WD. \\
    16 & References & & Literature references for the system. \\
    17 & WISE1 & mag & IR magnitude in WISE1-band. \\
    18 & e$\textunderscore$WISE1 & mag & Corresponding magnitude error in WISE1-band. \\
    19 & WISE2 & mag & IR magnitude in WISE2-band. \\
    20 & e$\textunderscore$WISE2 & mag & Corresponding magnitude error in WISE2-band. \\
    21 & WISE3 & mag & IR magnitude in WISE3-band. \\
    22 & e$\textunderscore$WISE3 & mag & Corresponding magnitude error in WISE3-band. \\
    23 & WISE4 & mag & IR magnitude in WISE4-band. \\
    24 & e$\textunderscore$WISE4 & mag & Corresponding magnitude error in WISE4-band. \\
    25 & 2MASSJ & mag & IR magnitude in 2MASS J-band. \\
    26 & e$\textunderscore$2MASSJ & mag & Corresponding magnitude error in 2MASS J-band. \\
    27 & 2MASSH & mag & IR magnitude in 2MASS H-band. \\
    28 & e$\textunderscore$2MASSH & mag & Corresponding magnitude error in 2MASS H-band. \\
    29 & 2MASSK & mag & IR magnitude in 2MASS K-band. \\
    30 & e$\textunderscore$2MASSK & mag & Corresponding magnitude error in 2MASS K-band. \\
    31 & UKIDSSY & mag & IR magnitude in UKIDSS Y-band. \\
    32 & e$\textunderscore$UKIDSSY & mag & Corresponding magnitude error in UKIDSS Y-band. \\
    33 & UKIDSSJ & mag & IR magnitude in UKIDSS J-band. \\
    34 & e$\textunderscore$UKIDSSJ & mag & Corresponding magnitude error in UKIDSS J-band. \\
    35 & UKIDSSH & mag & IR magnitude in UKIDSS H-band. \\
    36 & e$\textunderscore$UKIDSSH & mag & Corresponding magnitude error in UKIDSS H-band. \\
    37 & UKIDSSK & mag & IR magnitude in UKIDSS K-band. \\
    38 & e$\textunderscore$UKIDSSK & mag & Corresponding magnitude error in UKIDSS K-band. \\
    39 & VHSY & mag & IR magnitude in VHS Y-band. \\
    40 & e$\textunderscore$VHSY & mag & Corresponding magnitude error in VHS Y-band. \\
    41 & VHSJ & mag & IR magnitude in VHS J-band. \\
    42 & e$\textunderscore$VHSJ & mag & Corresponding magnitude error in VHS J-band. \\
    43 & VHSH & mag & IR magnitude in VHS H-band. \\
    44 & e$\textunderscore$VHSH & mag & Corresponding magnitude error in VHS H-band. \\
    45 & VHSKs & mag & IR magnitude in VHS Ks-band. \\
    46 & e$\textunderscore$VHSKs & mag & Corresponding magnitude error in VHS Ks-band. \\
    47 & GaiaG & mag & Optical magnitude in {\it Gaia} G-band. \\
    48 & e$\textunderscore$GaiaG & mag & Corresponding magnitude error in {\it Gaia} G-band. \\
    49 & GaiaBP & mag & Optical magnitude in {\it Gaia} BP-band. \\
    50 & e$\textunderscore$GaiaBP & mag & Corresponding magnitude error in {\it Gaia} BP-band. \\
    51 & GaiaRP & mag & Optical magnitude in {\it Gaia} RP-band. \\
    52 & e$\textunderscore$GaiaRP & mag & Corresponding magnitude error in {\it Gaia} RP-band. \\
    53 & SDSSu & mag & Optical magnitude in SDSS u-band. \\
    54 & e$\textunderscore$SDSSu & mag & Corresponding magnitude error in SDSS u-band. \\
    55 & SDSSg & mag & Optical magnitude in SDSS g-band. \\
    56 & e$\textunderscore$SDSSg & mag & Corresponding magnitude error in SDSS g-band. \\
    57 & SDSSr & mag & Optical magnitude in SDSS r-band. \\
    58 & e$\textunderscore$SDSSr & mag & Corresponding magnitude error in SDSS r-band. \\
    59 & SDSSi & mag & Optical magnitude in SDSS i-band. \\
    60 & e$\textunderscore$SDSSi & mag & Corresponding magnitude error in SDSS i-band. \\
    61 & SDSSz & mag & Optical magnitude in SDSS z-band. \\
    62 & e$\textunderscore$SDSSz & mag & Corresponding magnitude error in SDSS z-band. \\
    63 & GALEXFUV & mag & UV magnitude in GALEX FUV-band. \\
    64 & e$\textunderscore$GALEXFUV & mag & Corresponding magnitude error in GALEX FUV-band. \\
    65 & GALEXNUV & mag & UV magnitude in GALEX NUV-band. \\
    66 & e$\textunderscore$GALEXNUV & mag & Corresponding magnitude error in GALEX NUV-band. \\
    %\noalign{\smallskip}
    \hline
    %\noalign{\smallskip}

 %   \multicolumn{8}{l}{\footnotesize {\it References.} (1) \cite{stelzer2017}, (2) \cite{burleigh2006}, (3) \cite{munoz2023},}\\
%    \multicolumn{8}{l}{\footnotesize (4) \cite{kruckow2021}, (5) \cite{farihi2008},(6) \cite{breedt}}\\
\end{longtable}

\begin{table}[H]
    \centering
    \begin{threeparttable}
    \caption{Shortened version of our period-bounce candidates catalog showing relevant properties of the systems.}
    \label{PBtable}
    \begin{tabular}{c c cc c ccc c} 
    \hline
    \noalign{\smallskip}
    System & $P_{\rm orb}$ & M$_{\rm donor}$ & SpT$_{\rm donor}$ & Distance &  WD T$_{\rm eff}$ &  M$_{\rm WD}$ & WD type & References  \\
    & [h] & $\rm[M_{\odot}]$ & & [pc] & [K] & $\rm [M_{\odot}]$ &  & \\
    \noalign{\smallskip}
    \hline
    \noalign{\smallskip}
    V379\,Vir & 1.47 & 0.050 & L8 & 153 &  10000 & 0.64 & Magnetic & 1, 2, 3, 4, 5 \\
    SDSS 1514 & 1.47 & 0.082 & L3 & 182 & 10000 & 0.80 & Magnetic & 3, 6 \\
    SDSS 1250 & 1.44 & 0.077 & M8 & 131 & 10000 & 0.64 & Magnetic  & 6, 3 \\
    EF Eri & 1.35 & 0.083 & Not late & 160 & 10000 & 0.90 & Magnetic & 7, 8 \\
    SDSS 1057 & 1.51 & 0.044 & L5 & 355 & 11500 & 0.83 & Non-magnetic & 9, 10 \\
    \vdots & \vdots & \vdots & \vdots & \vdots & \vdots & \vdots & \vdots &\vdots \\
    \noalign{\smallskip}
    \hline
    \end{tabular}
    \begin{tablenotes}
        \small
        \item {\it References.} (1) \cite{stelzer2017}, (2) \cite{burleigh2006}, (3) \cite{munoz2023}, (4) \cite{kruckow2021}, (5) \cite{farihi2008}, (6) \cite{breedt}, (7) \cite{schwope2010}, (8) \cite{howell2006}, (9) \cite{mcallister2017}, (10) \cite{echevarria2023}.
    \end{tablenotes}
  \end{threeparttable}
\end{table}

\twocolumn
\section{Examples for SED types}\label{appen:SED}

We define 3 typical cases for SEDs of pre-bounce CVs, CVs at the period minimum and post-bounce CVs with the following criteria: excess setting in at wavelenghts shorter than 12483$\AA$ ($J$-band) for CVs before the period-bounce, excess setting in at wavelenghts between 12483$\AA$ and 22010$\AA$ ($K$-band) for CVs around the period-bounce, and excess setting in at wavelenghts longer than 22010$\AA$ for CVs significantly after the period-bounce. 

\begin{figure}
    \centering
    %trim={<left> <lower> <right> <upper>}
    \includegraphics[width=\columnwidth,trim=0.1cm 0.5cm 1.1cm 1.2cm,clip]{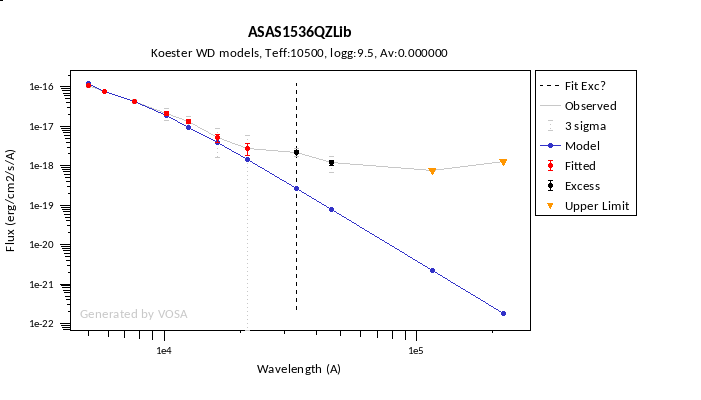}
    \caption{WD fit to the SED of QZ Lib. This system is a confirmed period-bouncer with a spectroscopic detected brown dwarf donor. VOSA marks the excess starting at WISE 1 band.}
    \label{SEDqzlib}
\end{figure}

\begin{figure}
    \centering
    %trim={<left> <lower> <right> <upper>}
    \includegraphics[width=\columnwidth,trim=0.1cm 0.5cm 1.1cm 1.2cm,clip]{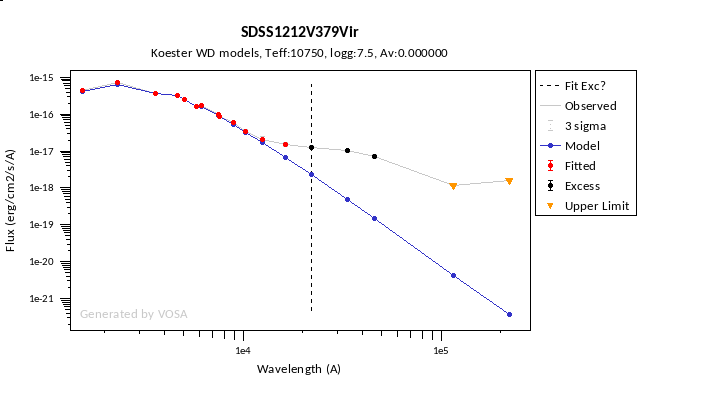}
    \caption{WD fit to the SED of V379 Vir. This system is a confirmed period-bouncer with a spectroscopic detected L8 donor. VOSA marks the excess starting at $K$-band.}
    \label{SED1212}
\end{figure}

\begin{figure}
    \centering
    %trim={<left> <lower> <right> <upper>}
    \includegraphics[width=\columnwidth,trim=0.1cm 0.5cm 1.1cm 1.2cm,clip]{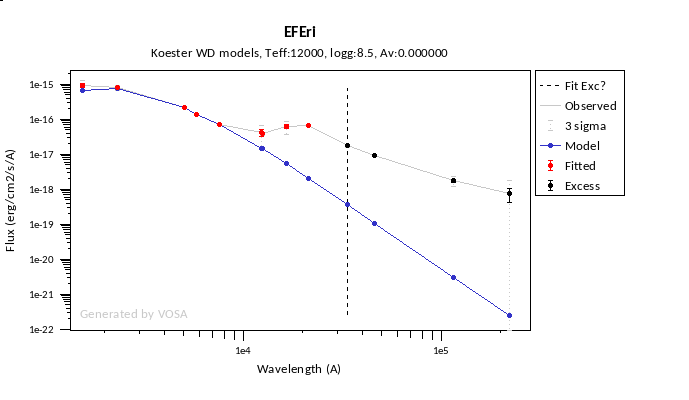}
    \caption{WD fit to the SED of EF Eri. This system has been proven to not be period-bouncer. VOSA marks the excess starting at WISE 1 band, but it is clear to us that it starts at $J$-band.}
    \label{SEDeferi}
\end{figure}

To determine the onset wavelength of the excess in the SED we used VOSA, were we fitted a WD model \citep{koester2010} to the SED of each period-bounce candidate in our literature catalog initially giving as input a range of $\pm$1000\,K around the WD temperature found in the literature. When necessary, we adjusted the temperature range to obtain a better fit specifically for the GALEX points considering that it is in the UV bands where we get the best constraints on the WD. The final WD temperature and log(g) values used in the fit are reported in the top part of each figure. We did not consider extinction. VOSA suggests a point for the beginning of the excess (vertical dashed line in Figs.~\ref{SEDqzlib}, \ref{SED1212} and \ref{SEDeferi}) which, in most cases, we used to assign the corresponding points in our scorecard. There were a few cases (see Fig.~\ref{SEDeferi}) where it was clear that the excess starts at a shorter wavelength, in which case we selected ourselves the shorter band as the point of beginning of the excess.

%----------------------------------------------------------------- 
%-----------------------------------------------------------------

%----------------------------------------------------------------- 
%-----------------------------------------------------------------

%----------------------------------------------------------------- 
%-----------------------------------------------------------------

\end{appendix}

\end{document}